\documentclass[twocolumn]{aastex62}

\usepackage{amsmath}
\usepackage{graphicx}
\usepackage{dcolumn}
\usepackage{bm}
\usepackage{xcolor}
\usepackage{fvextra} 
\usepackage{codebox}

\graphicspath{{./}{figures/}}

\begin{document}

\title{Laboratory measurements of energy partitioning and anomalous electron heating in magnetized,\\ perpendicular collisionless shocks}

\correspondingauthor{Vicente Valenzuela-Villaseca}
\email{v.valenzuela@princeton.edu, valenzuelavi1@llnl.gov,\\ valenzuela@psfc.mit.edu}

\author[0000-0002-0786-7307]{V. Valenzuela-Villaseca}
\altaffiliation{Lawrence Postdoctoral Fellow}
\affiliation{Department of Astrophysical Sciences, Princeton University, Princeton, New Jersey 08544, USA.}
\affiliation{Lawrence Livermore National Laboratory, Livermore, California 94550, USA.}
\affiliation{Plasma Science and Fusion Center, Massachusetts Institute of Technology, Cambridge, Massachusetts 02139, USA}

\author{S. Totorica}
\affiliation{Department of Astrophysical Sciences, Princeton University, Princeton, New Jersey 08544, USA.}
\affiliation{Department of Astro-fusion Plasma Physics (AFP), Headquarters for Co-Creation Strategy, National Institutes of Natural Sciences, Tokyo 105-0001, Japan}

\author{J. Griff-McMahon}
\affiliation{Department of Astrophysical Sciences, Princeton University, Princeton, New Jersey 08544, USA.}
\affiliation{Princeton Plasma Physics Laboratory, Princeton, New Jersey 08540, USA.}

\author{L.-J. Chen}
\affiliation{Geospace Physics Laboratory, NASA Goddard Space Flight Center, Greenbelt, Maryland 20771, USA.}

\author{S. Malko}
\affiliation{Princeton Plasma Physics Laboratory, Princeton, New Jersey 08540, USA.}

\author{P. V. Heuer}
\affiliation{Laboratory for Laser Energetics, University of Rochester, Rochester, New York 14623, USA}

\author{P. Pongkitiwanichakul}
\affiliation{Department of Physics, Kasetsart University, Lat Yao, Chatuchak, Bangkok 10900, Thailand.}

\author{W. Fox}
\affiliation{Department of Astrophysical Sciences, Princeton University, Princeton, New Jersey 08544, USA.}
\affiliation{Princeton Plasma Physics Laboratory, Princeton, New Jersey 08540, USA.}
\affiliation{Department of Physics, University of Maryland, College Park, Maryland 20742, USA.}

\author{D. B. Schaeffer}
\affiliation{Department of Physics and Astronomy, University of California Los Angeles, Los Angeles, California 90095, USA}

\begin{abstract}

We present laboratory results on energy partitioning from supercritical, magnetized collisionless shock experiments ($\rm{M_A} \sim 8$, $\rm{M_{ms}}\sim 4$). We report the first observation of fully-developed laboratory shocks that evolve for more than seven upstream ion gyroperiods and have a downstream region that extends more than four shocked ion gyroperiods. We use Thomson scattering to measure electron and ion temperatures, plasma density, and flow speeds.  We directly measure a compression ratio of $3.6\pm0.3$, consistent with shock jump conditions.  A foot ahead of the shock exhibits super-adiabatic electron and ion heating.  The downstream electron temperature exceeds adiabatic and collisional electron-ion heating by $\approx 30\%$, implying significant collisionless anomalous electron heating. We find a downstream electron-ion temperature ratio $T_e^{(d)}/T_i^{(d)} = 0.8 \pm 0.3$, consistent with spacecraft observations but outside the range predicted by theory and numerical simulations. 

\end{abstract}



\section{Introduction} \label{sec:intro}

Collisionless shocks are common structures throughout the universe, found in planetary magnetospheres \citep{Smith1975,Smith1980,Sulaiman2015,Lefebvre2007, wilson+16b,johlander+21, Lalti+22,Schwartz2022}, supernova remnants \citep{Cargill1987,Spicer1989,Bamba2003,rothenflug+04,gamil+08,SN1006HESS,bocchino+11,giuffrida+22}, and when galaxies merge inside a cluster \citep{willson70, fujita+01,govoni+04,vanweeren+10,brunetti+14,lindner+14}. In a collisionless shock, the ion-ion Coulomb mean-free-path $\lambda^{i \backslash i}_{\text{mfp}}$ greatly exceeds the density gradient scale-length $L_n \equiv n/|\nabla n|$, implying that collective electromagnetic fields in the plasma mediate interactions, not collisional viscosity \citep{Balogh2013,Burgess2015,Marcowith2016}.  How the inflowing kinetic energy is partitioned by the shock between ions and electrons is a crucial parameter controlling the thermodynamics, radiation, and particle acceleration in collisionless shocks.  While magnetohydrodynamic (MHD) theory dictates how bulk properties such as the temperature change from the far upstream to far downstream across a shock, dynamics through the shock are inherently nonlinear and kinetic, with no generally accepted theory of particle heating and energy partitioning in collisionless shocks \citep{Balikhin1993,Lembge2003,Schwartz2011,Schwartz2022,Stasiewicz2023}.  

MHD theory predicts that the energy balance across a shock is dominated by ions, since the ion kinetic energy dominates in the shock upstream.  A common closure in MHD theory is that electrons heat up through adiabatic compression, which typically leads to ion temperatures much greater than electron temperatures downstream of the shock.  This is often in agreement with kinetic simulations (e.g., see \citealt{guo_electron_2017, Tran2020}).  \textit{In situ} spacecraft measurements of planetary bow shocks in the solar system do show temperature differences between electrons and ions, typically parameterized by the downstream electron-ion temperature ratio $T_e/T_i$, but they also observe energy partitioning that is much closer to equipartition than expected from the electron/ion mass ratio \citep{Schwartz1988,Schwartz2011}.  When combined with observations of collisionless shocks in supernova remnants, these measurements suggest that energy partitioning trends nonlinearly with the shock Mach number (first decreasing, then increasing again at intermediate Mach numbers, before asymptoting at high Mach numbers) \citep{Tran2020,Raymond2023}, but the exact dependence remains an outstanding question.

Satellite observations can provide empirical tests of candidate theories, but bow-shock crossings in the heliosphere show significant variance in electron heating, especially for intermediate magnetosonic regions ($1\lesssim \rm{M_{ms}}\lesssim 10$) \citep{Schwartz2022,Schwartz1988,Chen2018}, leading to highly noisy statistics from which it is challenging to extract trends. Remote observations of supernova remnant shocks are even more limited due to challenges in measurements of electrons, neutral populations, and upstream parameters \citep{Raymond2023}.  As a result, there have been several efforts to study shock heating through numerical simulations using particle-in-cell (PIC) codes, which can capture the electron dynamics needed to diagnose intrinsically kinetic processes.  These studies have captured the overall trends of energy partitioning as a function of Mach number \citep{Bohdan2020, Tran2020, Morris2023, Vanthieghem2024, Tran2024}, but they have been unable to reproduce the full range of observed temperature ratios.  In part, this is due to the computationally prohibitive expense of trying to capture the large range of scales required for realistic 3D system sizes and electron-ion mass ratios.

Controlled laboratory experiments can provide independent measurements of energy partition in collisionless shocks that can complement theory, astronomical and heliospheric data, and numerical simulations by providing absolute measurements of the temperature ratio.  The primary experimental approach is to use high-powered lasers to create dimensionally-scaled collisionless shocks in the laboratory \citep{Drake2000}.  Laser-driven experiments in initially unmagnetized plasmas created Weibel-mediated collisionless shocks and measured a downstream temperature ratio $T_e/T_i\approx0.3$ \citep{Fiuza2020}, consistent with numerical simulations at very high Mach number.  In pre-magnetized plasmas, experiments have utilized laser-driven magnetic pistons \citep{Schaeffer2014,Bondarenko2017} to create quasi-perpendicular collisionless shocks with moderate Mach numbers ($1\lesssim M_{ms}\lesssim 10$) \citep{Schaeffer2017,Schaeffer2019,Matsukiyo2022}.  These previous experiments measured downstream temperatures, but the shocks were confined to short lifetimes of only a few ion gyrotimes, and so the shock downstream could not fully develop and separate from the piston \citep{Schaeffer2020}. Other experiments \cite{Yao2021} with lower driving laser energies observed shock velocities that depended strongly on time (i.e. the shock appreciably decelerated in the experimental time frame), leading to large uncertainty in the upstream parameters needed to evaluate shock jump conditions and energy partitioning.

In this paper, we report the experimental observation of a fully-developed quasi-perpendicular collisionless shock.  The shock is supercritical ($\rm{M_A} \sim 8$, $\rm{M_{ms}} \sim 4$, $\beta \sim 3$) and observed over seven gyrotimes, sufficient to develop an extended downstream region extending over four shocked ion gyroperiods and well separated from the driving piston.  The shock structure is probed \textit{in situ} as it passes through a stationary optical Thomson scattering collection volume, scanning from the unperturbed upstream (denoted by $\Box^{(u)}$) through the shock into the downstream (denoted by $\Box^{(d)}$) and piston region. We observe a shock ramp width consistent with the shock-crossing time of a shocked proton, and directly measure the compression ratio $R = n_e^{(d)}/n_e^{(u)} = 3.6 \pm 0.3$ and super-adiabatic electron heating $T_e^{(d)}/T_e^{(u)} = 3.0 \pm 0.4 > R^{2/3}=2.3\pm0.1$. Ion temperatures are found to increase from upstream $T_i^{(u)} = 15 \pm 4$ eV to downstream $T_i^{(d)} = 460 \pm 170$ eV, implying a temperature ratio $T_e^{(d)}/T_i^{(d)} = 0.8 \pm 0.3$, outside the bounds of magnetohydrodynamics (MHD) theory and previous simulations \citep{Tran2020}. The data also show electron and ion super-adiabatic pre-heating in a foot region ahead of the shock. 

Our article is organized as follows: \autoref{sec:exp setup} describes the laser-driven experiment and the plasma diagnostic utilized in this study, namely optical Thomson scattering. The results are presented in \autoref{sec:results}, showing the evolution of plasma parameters in shock experiments, together with two null experiments that remove either the upstream medium or the background magnetic field. \autoref{sec:discussion} discusses the downstream ion temperature and classical electron heating mechanisms (i.e., adiabatic compression and electron-ion collisional equilibration) and shows that the downstream electron temperature cannot be explained by these two effects, implying that the electron heating mechanism must be collisionless. Additionally, we discuss energy partition in the context of data from heliospheric and supernova remnant observations, MHD theory, and kinetic simulations. We present our conclusions and mention avenues of future studies in \autoref{sec:conclusion}. Details about the data analysis pipeline and analytical calculation (particularly regarding multi-ion population plasmas) are presented in the \hyperref[sec:appendix]{Appendix}.

\section{Experimental setup and methods} \label{sec:exp setup}

\begin{figure}
    \centering
    \includegraphics[width=0.8\linewidth]{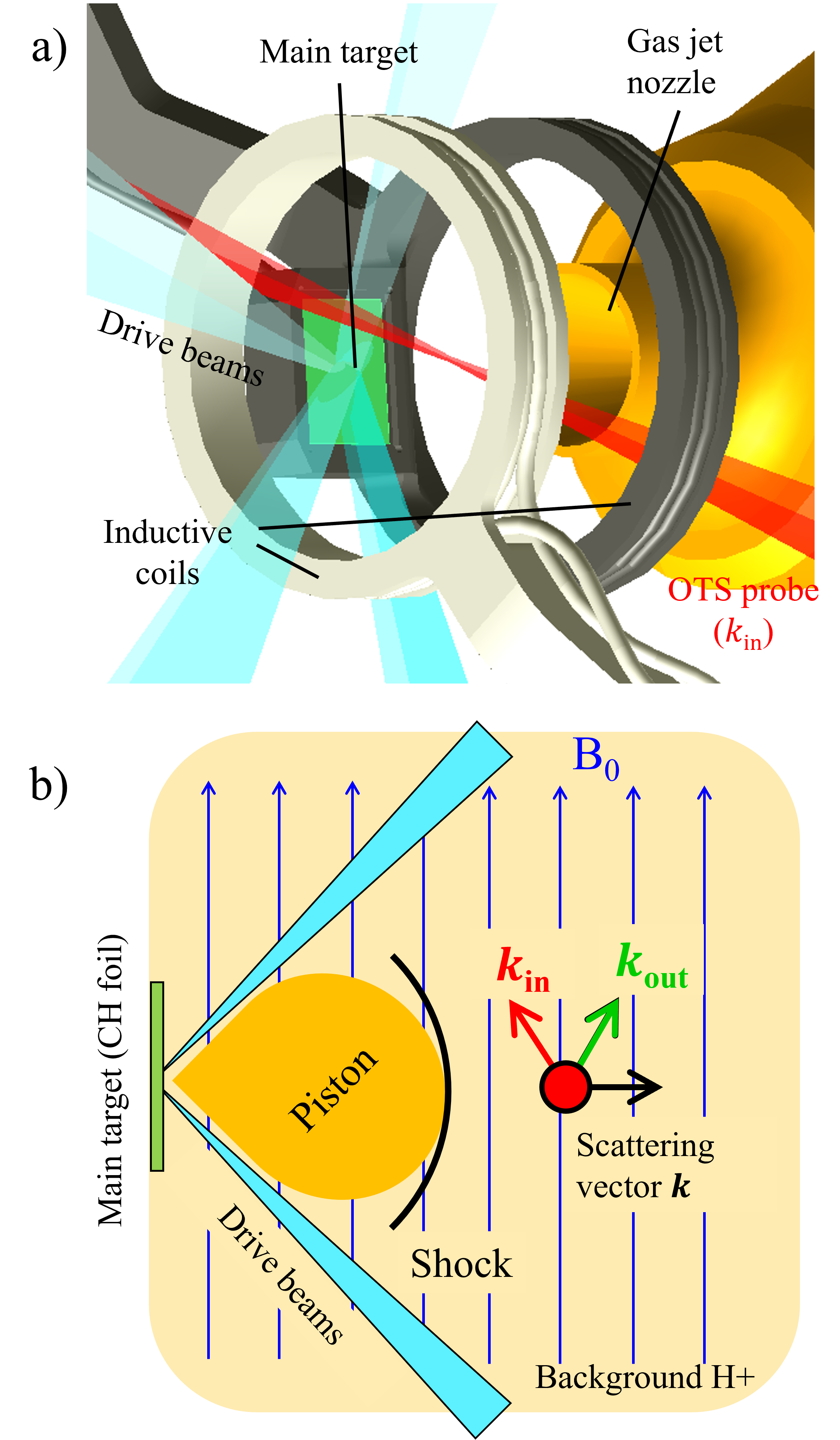}
    \caption{Experimental setup and data from experiments with background density $n_{e}^{(u)} = 5 \times 10^{18}$ cm$^{-3}$ and magnetic field $B^{(u)} = 10.5$ T. a) 3-D model of experimental setup. Drive laser beams are shown in blue and optical Thomson scattering probe in red. b) Schematic of fundamental experimental operation. }
    \label{fig:exp setup}
\end{figure}

The experiments were conducted on the OMEGA laser \citep{boehly+95} at the Laboratory for Laser Energetics. As shown in \autoref{fig:exp setup}a, the experiment consists of irradiating a flat plastic (CH) foil by four laser beams  (351 nm wavelength, 375 J/beam, 2-ns square pulses, $10^{14}$ W/cm$^2$ intensity) fired simultaneously with a super-Gaussian distributed phase plate spatial profile. The target is positioned between two Helmholtz coils driven by the Magneto-Inertial Fusion Electrical Discharge System (MIFEDS) \citep{Fiksel2015}. The coils, two parallel $8$ mm-radius double loops separated by $8$ mm, create a upstream uniform magnetic field $B^{(u)} = 10.5$ T. A  Mach $5$, $6$-mm-diameter gas jet nozzle, parallel to the target surface, supplied the upstream plasma with electron density $n_e^{(u)} \sim 5 \times 10^{18}$ cm$^{-3}$. 

As shown schematically in \autoref{fig:exp setup}b, laser irradiation drives a supersonic plasma plume that acts as a piston \citep{Schaeffer2017} to compress and accelerate the upstream magnetized medium, launching a shockwave. In addition, the laser-solid interaction releases an intense X-ray burst that pre-ionizes the background hydrogen, creating a stationary electron-proton plasma upstream of the shock. As shown below, the shock is observed to propagate at approximately $v_{\text{sh}}\sim650$ km/s, which is sufficient to establish
collisionless conditions. The Coulomb mean-free-path for protons moving at that velocity through a stationary proton background $\lambda^{i \backslash i}_{\text{mfp}} = 5\times 10^{7} A~(v\text{ [km/s]})^4 / \bar{Z}^3 n_e \text{ [cm$^{-3}$]} \sim 20$ mm 
exceeds the system size, implying the ion dynamics are collisionless ($A=1$ is the atomic weight and $\bar{Z}=1$ is the effective charge state). 

Based on this shock velocity we also evaluate an Alfv\'enic Mach number $\rm{M_A}$$\equiv v_{\text{sh}}/V_{A}^{(u)} \sim 8$ and magnetosonic Mach number $\rm{M_{ms}}$$\equiv v_{\text{sh}}/ V_{\rm ms}^{(u)}\sim 4$, where $V_{A}^{(u)} = B^{(u)}/\sqrt{\mu_0 m_i n_{i}^{(u)}}$ is the upstream Alfv\'en velocity ($\mu_0$ is the vacuum permeability, $m_i$ is the proton mass, and $n_{i}^{(u)}$ the upstream ion density), $c_s^{(u)} = \left[\gamma \bar{Z} k_BT_e^{(u)}/m_i\right]^{1/2}$ the upstream ion sound speed ($\gamma=5/3$ is the adiabatic index and $k_B$ is Boltzmann's constant), and 
$V^{(u)}_{\rm{ms}}$$=\left[\left(V_{A}^{(u)}\right)^2 + \left(c_{s}^{(u)}\right)^2\right]^{1/2}$ the upstream magnetosonic speed. 

The plasma is probed using optical Thomson Scattering (OTS) \citep{Katz2012}, which provides 
a localized, time-resolved measurement of the plasma parameters by analyzing the shape of the 
scattered light spectrum.  In particular, when the particle distributions are Maxwellian, the underlying plasma parameters are obtained by best fits to theoretical Thomson spectra, from which measurements of $n_e$, $\bar{Z}$, $T_e$, $T_i$, and ion flow velocity $u$ are obtained. 

The probe beam ($\lambda_i=526.5$ nm incident wavelength, 150 J, 3.7-ns square pulse, 100 $\mu$m DPP spatial profile) is focused $5$ mm from the target foil and starts $6.5$ ns after the target foil is irradiated.  This corresponds to $\approx8~r_i^{(u)}$ and $\approx7~\left[\omega_{ci}^{(u)}\right]^{-1}$, where the upstream ion gyroradius $r_i^{(u)}=v_{sh}/\omega_{ci}^{(u)}$, and $\omega_{ci}^{(u)}$ is the upstream ion frequency. The probe direction defines the incident wave-vector $\mathbf{k_{in}}$ in the scattering interaction. The scattered light was collected at an angle $\theta=63^{\circ}$ from $\mathbf{k_{in}}$, defining the scattered vector $\mathbf{k_{out}}$. The scattered spectrum is determined by Doppler shifts parallel to the scattering vector $\mathbf{k} \equiv \mathbf{k_{out}} - \mathbf{k_{in}}$, and in these experiments we set up $\mathbf{k_{in}}$ and $\mathbf{k_{out}}$ so that $\mathbf{k}$ was oriented parallel to the shock propagation and perpendicular to the magnetic field (\autoref{fig:exp setup}b).  We emphasize that the diagnostic directly probes velocity and temperature components parallel to $\mathbf{k}$. The scattered spectrum is described in terms of the angular frequency $\omega=\omega_s-\omega_i$, where $\omega_i$ and $\omega_s$ are the incident (laser) and scattered angular frequency, respectively.  For convenience, this is converted to wavelength $\lambda$ when plotted. The experiments are in the collective regime, with a scattering parameter $\alpha = 1/k\lambda_{De} \approx 2$, where $k=|\mathbf{k}|$ and $\lambda_{De}=\sqrt{\epsilon_0 k_B T_e/en_e }$ is the electron Debye length ($\epsilon_0$ is the electric permittivity of free space, and $e$ the fundamental charge). The scattered spectrum is described by the dynamic form factor \citep{Froula2011}
\begin{equation}\label{eq:Skw}
    S(k,\omega) = \frac{2\pi}{k} \left| 1 - \frac{\chi_e}{\varepsilon} \right|^2 f_{e0} + \frac{2\pi}{k}\sum_s \frac{\bar{Z}_s n_s}{n_e} \left| \frac{\chi_e}{\varepsilon} \right|^2 f_{s0},
\end{equation}
where $\varepsilon = 1 + \chi_e + \sum_s \chi_s$ is the plasma dielectric function, $f_{s0}$ and $f_{e0}$ are velocity distribution functions (VDFs) of the ion species $s$ and electrons, respectively, $\chi_s$ and $\chi_e$ are the susceptibilities of ions and electrons, and $n_s$ is the number density of ion species $s$. The first term in equation (\ref{eq:Skw}) corresponds to the electron-plasma wave (EPW) feature and the second to the ion-acoustic wave (IAW) feature. The scattered power spectrum $P(k,\omega) \propto S(k,\omega)(1+2\omega/\omega_i)$ was fit using a modified routine of \texttt{PlasmaPy} \citep{plasmapy2024} to include the $(1+2\omega/\omega_i)$ relativistic correction (see \autoref{apx:tswift}). Uncertainties were estimated by manually varying each parameter until the model no longer fit the data (see e.g. \cite{Suttle2021,Valenzuela-Villaseca2023}).

\newpage
\section{Results} \label{sec:results}

\begin{figure}
    \centering
    \includegraphics[width=1.02\linewidth]{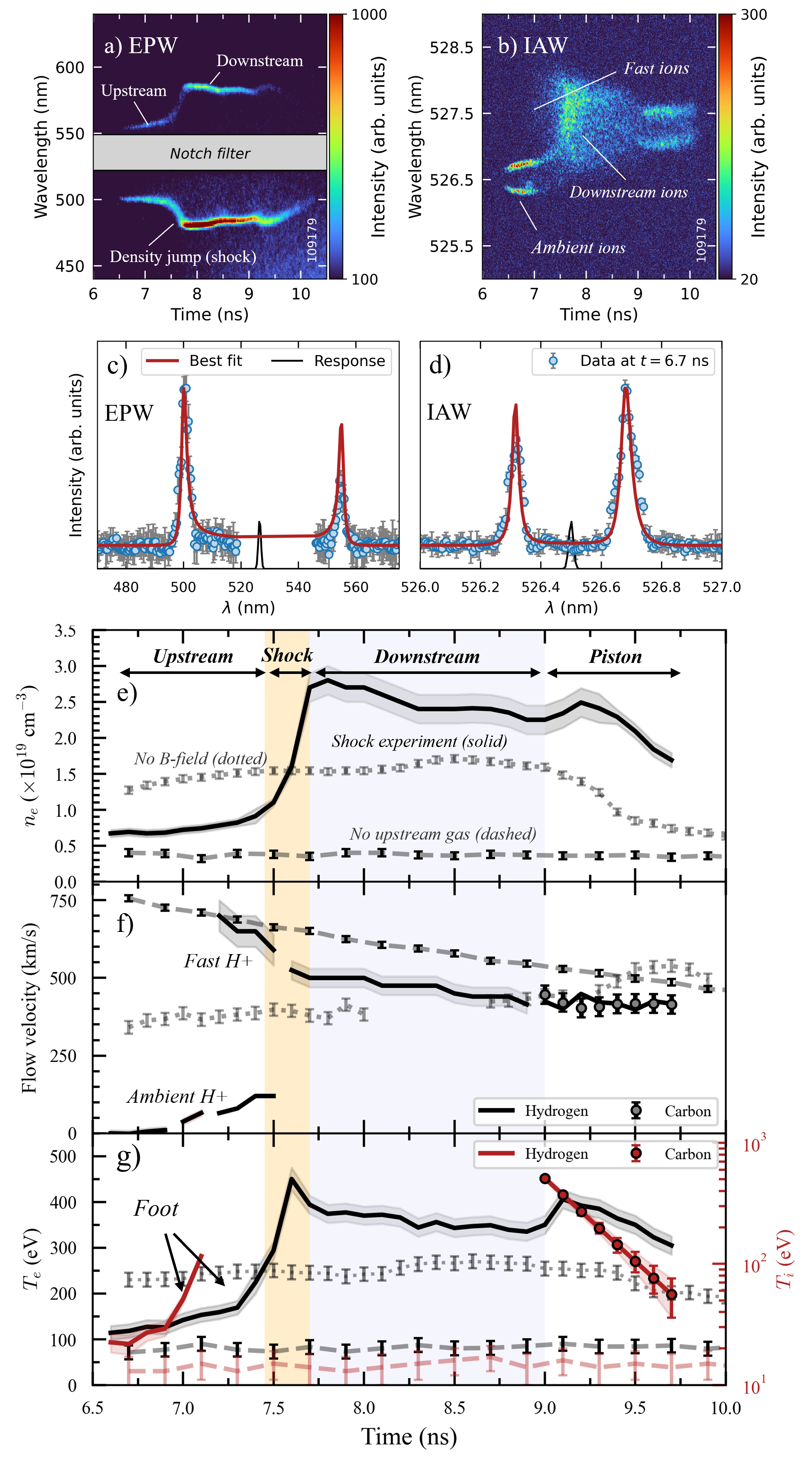}
    \caption{OTS spectra and evolution of measured plasma parameters. a) EPW CCD spectral image of the shock experiment. A notch filter is used to remove the IAW feature. b) IAW CCD spectral image of the shock experiment. c) EPW fit at 6.9 ns. d) IAW fit. The instrument response function is shown as the black curve in (c) and (d). e) Electron density. f) Ion flow velocity. In the case of the shock experiments, circle markers indicate the flow velocity of carbon ions, whereas solid line denotes hydrogen. g) Left axis: Electron temperature (black, grey). Right axis: Ion temperature (red). In panels (e) to (g), plasma parameters measured in shock experiments (i.e. piston propagating through premagnetized upstream hydrogen) are presented in solid lines, whereas results without upstream gas or magnetic field are denoted with dashed or dotted lines, respectively}
    \label{fig:shock evolution}
\end{figure}

\autoref{fig:shock evolution}a and b) present the time- and wavelength-resolved raw spectral data for an experiment observing the transit of a collisionless shock past the Thomson scattering volume. We first discuss the qualitative features of the data before discussing detailed fitting below.

The EPW data (Fig.~\ref{fig:shock evolution}a) shows a strong time dependence of the separation of the red- and blue- shifted resonances, where $\Delta \omega \approx \pm \omega_{pe} \propto n_e^{1/2}$.
Meanwhile, the IAW spectrum (Fig.~\ref{fig:shock evolution}b) provides detailed information about the ions.  We immediately recognize that the ion spectrum strongly redshifts and broadens
at the shock transit time.  Since ion centroid $\propto \mathbf{k}\cdot \mathbf{u}$, this indicates a rapid velocity jump of the ions near the shock.

These data indicate the passage of a shock past the scattering
volume, including a rapid jump in the plasma density (via the EPW) and acceleration of the ions (via the IAW) relative to the lab frame, as the measurement first observes the upstream, through the shock, the downstream, and finally the piston.  We indicate the 
timing on Fig. 2(a,b) as follows: From $t = 6.5$ ns to $7.5$ ns, the diagnostic probes the upstream, where the plasma is low density and stationary. At $t=7.5$ ns, the EPW resonances rapidly separate indicating a steep density jump; this is the shock front. Simultaneously, the IAW spectrum broadens (losing the double-peak resonance) and redshifts going into the downstream. In addition, slightly before the shock at $t=7$ ns, we also note that a secondary Doppler-shifted IAW feature from fast ions emerges, potentially indicative of ions reflected from the shock front. For $t\gtrsim9$ ns, the double-peaked IAW spectrum re-emerges, indicating the piston plasma.

We study the shock structure using spectral lineouts averaged over $0.1$ ns, corresponding to a spatial resolution of $\approx65$ $\mu$m based on the shock velocity, which is comparable to the Thomson scattering volume and equivalent to $\approx0.5$ shocked ion gyroradii $r_{i}^{(d)}=u^{(d)}/\omega_{ci}^{(d)}$. When consistent with a Maxwellian VDF, equation (\ref{eq:Skw}) is fit to extract plasma parameters (see \autoref{fig:shock evolution}c,d for example spectra). We find that the EPW feature is Maxwellian throughout the experiment, whereas the IAW becomes non-Maxwellian in the downstream. Figure~\ref{fig:shock evolution}e-g shows the plasma evolution in shock and null experiments. We focus first on the shock experiments (solid lines), and return to the null experiments below. The upstream density $n_e^{(u)} = (6.7\pm0.5) \times 10^{18}$ cm$^{-3}$ jumps to a peaked value of $ (2.8\pm 0.1) \times 10^{19}$ cm$^{-3}$ as the shock passes through the probe at $t=7.6\pm 0.2$ ns. The time-of-flight indicates a shock speed $v_{sh} = 660 \pm 20$ km/s. After this initial overshoot, the density decreases to $n_e^{(d)} = (2.4\pm0.1) \times 10^{19}$ cm$^{-3}$, giving a compression ratio $R = 3.6\pm 0.3$, consistent with a fully developed shock and the MHD jump conditions.  From this, we assume a downstream compressed magnetic field of $B^{(d)} = R B^{(u)}=37.6\pm 3.2$ T, based on previous experiments we have done that have observed field compressions consistent with the jump conditions \citep{Schaeffer2019}.

The shock has a ramp time $\tau_{\text{ramp}} = 0.25$ ns.  This is approximately equal to $r_{i}^{(d)}/v_{sh}$, the shock-crossing time over one shocked proton gyroradius in the downstream field. We identify the upstream, shock, and downstream regions shown in \autoref{fig:shock evolution}e from these measurements. The ion flow velocity $u$ is measured either directly through the $S(k,\omega)$ for Maxwellian ions or through the spectral centroid redshift in the downstream. Figure~\ref{fig:shock evolution}f shows that the upstream is initially at rest. At $t=7.2$ ns, a secondary Doppler-shifted feature reveals a low density ion population streaming ahead of the shock, which might indicate the presence of reflected ions. However, the signal-to-noise ratio is too low to carry out a detailed analysis. Once the density jumps, the IAW spectrum is strongly shifted, indicating downstream flow with a characteristic velocity $u^{(d)}=475 \pm 30$ km/s. The shock velocity is related to the downstream ion flow speed and compression ratio as
\begin{equation}\label{eq:vsh}
    v_{sh} = \frac{R}{R-1} u^{(d)} = 660\pm50\, \rm{km/s},
\end{equation}
consistent with the observed shock time-of-flight and implying that the shock is quasi-steady-state (i.e. not accelerating or decelerating). 

The temperature evolution is shown in \autoref{fig:shock evolution}g. The upstream electron temperature $T_e^{(u)} = 115 \pm 15$ eV initially, far from the shock. The electrons heat up super-adiabatically ahead of the shock, overshooting to $450 \pm 25$ eV before relaxing to $T_e^{(d)} = 350 \pm 20$ eV in the downstream, where it remains approximately constant. 
We also find that, like the electrons, the upstream protons heat up from $T_i^{(u)} = 15 \pm 4 \, \rm{eV} \rightarrow 115 \pm 10$ eV ahead of the shock.

Finally, we can analyze the spectra to infer the ion fractional populations (H vs. C) at various times. While the upstream is initially entirely H ions (protons) from the gas jet, because of the plastic target foil, the total ion population can consist of both protons and carbon ions.  The H and C ion abundances are informed by the EPW electron temperature. The IAW peak separation is $\sim \bar{Z}T_e$, with effective ion charge state $\bar{Z} = f_H\bar{Z}_H + f_C\bar{Z}_C$, where $f_H$ and $f_C$ are the hydrogen and carbon fractions, respectively, and $\bar{Z}_H=1$ (for protons) and $\bar{Z}_C=6$ (for fully ionized carbon \footnote{Carbon is fully ionized for $T_e \geq 100$ eV at these densities \citep{Chung2005}}). In this regime, given $T_e$, a single value of $\bar{Z}$ is extracted from the IAW spectral fit, from which ion fractions are inferred. As expected, we find an electron-proton upstream plasma (no carbon) and a 50$\%$-50$\%$ proton-carbon mix in the piston ($t\geq 9$ ns). 

\begin{table*}
\small
\centering
\begin{tabular}{lcccc}
\hline
\hline
                                               &   Symbol      &   OMEGA Experiment    &   Earth's Bow Shock   &   Supernova Remnant Shock 1006 \\
                                                 &       &                       &   \citep{Schwartz2022} &   \citep{Bamba2003}            \\
\hline
System size& $L$                                        &	10 mm				&    100-1000 km        &   $10^{13}$ km        \\
Upstream magnetic field& $B^{(u)}$                       &	10.5 T              & $5\times10^{-5}$ G    &   $10^{-5}$ G			\\
Upstream ion density & $n_{i}^{(u)}$                      &	$6.7\times10^{18}$ cm$^{-3}$	&  5 cm$^{-3}$          &   0.5 cm$^{-3}$	    \\
Upstream ion inertial length & $d_{i}^{(u)}$              &	88 $\mu$m			&    100 km             &   300 km              \\ 
Upstream ion gyroperiod & $\left[\omega_{ci}^{(u)}\right]^{-1}$           &	1.0 ns				&   2 s                 &   10 s                \\
Shock speed & $v_{sh}$                                    &	660 km/s			&  400 km/s             &   2600 km/s	        \\
\hline
Upstream plasma beta & $\beta^{(u)}$                      &	3					&   1                   &   20                  \\
Alfv\'{e}nic Mach number & $\rm M_A$                          &	8					&   8                   &   84                  \\
Magnetosonic Mach number & $\rm M_{ms}$                       &   4                   &   6                   &   20                  \\
Electron magnetization & $\omega_{ce}^{(u)}/\nu_{ei}$     &	12				    &   $10^7$              &   $10^9$              \\
Ion collisionality & $\lambda_{ii}/d_{i}^{(u)}$           &	113					&   $\gg1$              &   $\gg1$              \\
Magnetic Reynolds number & \textbf{$\rm {Rm}$$=d_iv_{sh}/\eta$}                          &   87                  &   $10^9$              &   $10^{12}$           \\
\hline
\hline
\end{tabular}
\caption{Comparison of typical plasma and dimensionless parameters for laboratory and astrophysical collisionless shocks. The parameter $\eta$ is the Spitzer magnetic diffusivity.}
\label{tab:compare}
\end{table*}


It is valuable to compare the shock experiments to two null experiments (\autoref{fig:shock evolution}e-g, dashed and dotted lines) to confirm the collisionless nature of the shock. Crucially, in the absence of an upstream medium and/or a magnetic field, no density jump is observed, demonstrating that both are necessary to form the shock. Similarly, the plasma does not exhibit heating nor sudden velocity jumps in the null experiments. This further highlights that the magnetic field mediates the shock interaction, and Coulomb collisions are too inefficient to support significant energy and momentum dissipation. 

Several other observations from the null experiments also bear on the present results. We find a fast ion population streaming ahead of the shock at $\approx 700$ km/s, which is similar to the speed of the piston flowing through vacuum, as shown in \autoref{fig:shock evolution}f. Due to this similarity, it is possible that the fast ions observed in the shock experiment originate in the laser-target interaction, rather than reflected by the shock. More relevant to our study, we see a significant difference in electron temperatures in shock and null experiments [\autoref{fig:shock evolution}g]. The electrons in the piston have characteristic temperatures of $\approx 100$ eV, whereas when the upstream is present but the magnetic field is off, we observe electron temperatures of $\approx 250$ eV. When both upstream and magnetic field are present, the electrons reach temperatures of $\approx 400$ eV. Therefore, we conclude that the electron heating must be mediated by the shock (compression, collisional heating, and/or collisionless effects). We further analyze the pressure budget and electron heating in the discussion below.

\section{Discussion: energetics and heating analysis} \label{sec:discussion}

From our measurements we find that the experiments create dimensionally-scaled magnetized collisionless shocks directly relevant to quasi-perpendicular shocks observed around the Earth or in supernova remnants, as shown in Table~\ref{tab:compare}.  To investigate the energy partition across the shock, we now analyze the downstream energetics.

We first analyze the downstream ion temperature via a pressure balance analysis across the shock, supported by a scattered spectrum analysis, in order to measure the downstream energy partitioning $T_{e}^{(d)} / T_{i}^{(d)}$.  We then analyze the heating mechanisms for the electrons across the shock and find that the heating is anomalous and larger than can be explained by adiabatic
compression and electron-ion collisions.

\subsection{Inferring the downstream ion temperature}

\begin{figure}
    \centering
    \includegraphics[width=1\linewidth]{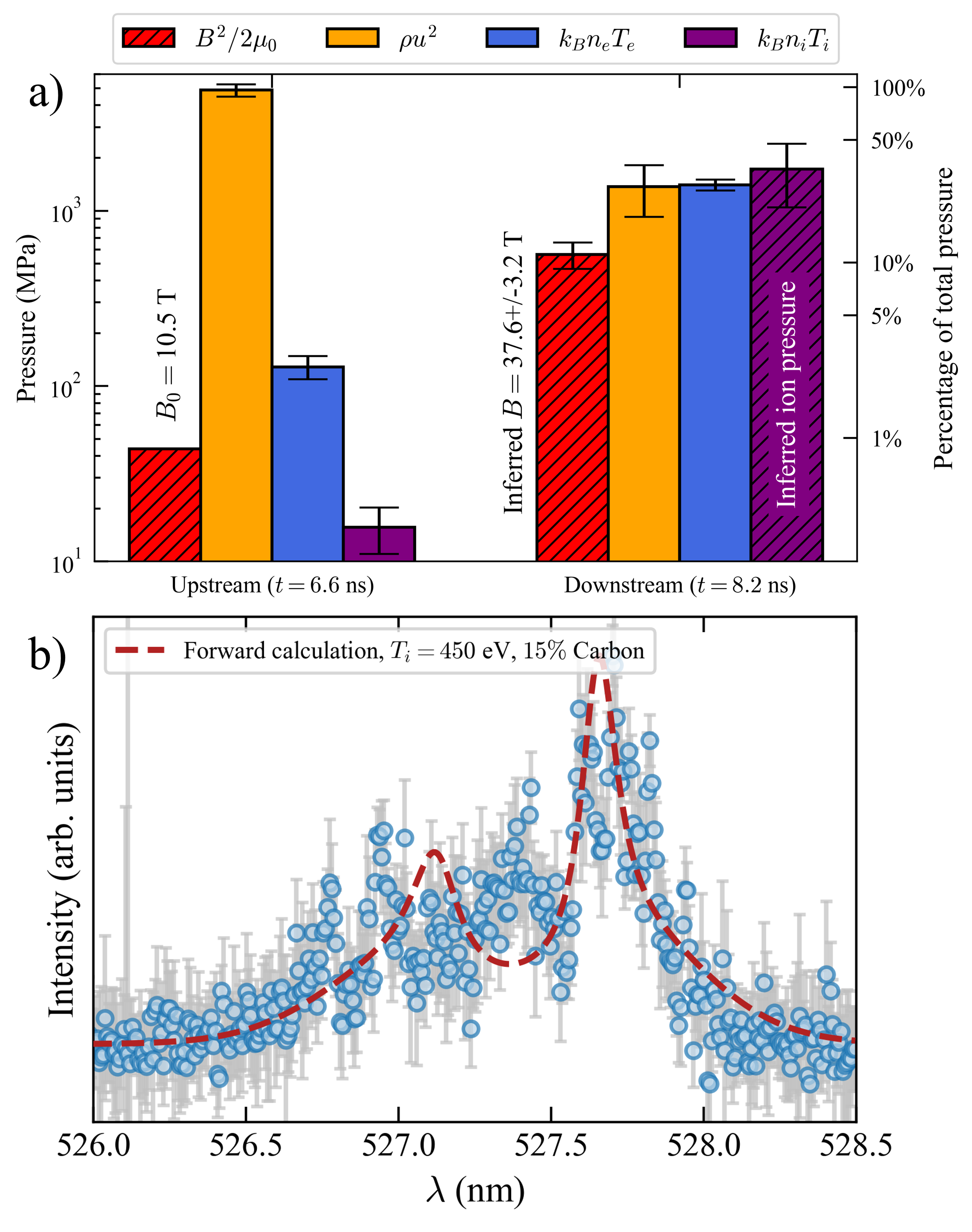}
    \caption{a) Upstream and downstream pressure components. Dashed bars denote inferred values. b) Upstream IAW spectrum at $t=8.2\,\rm{ns}$ compared to a forward model calculation at the specified conditions.} 
    \label{fig:pressures}
\end{figure}

We first analyze the ion temperature in the shock downstream. The IAW feature departs from Maxwellian VDFs there, preventing us from fitting the spectra and directly extracting $T_i^{(d)}$. Nevertheless, an estimation can be obtained by comparing the data with characteristic spectra for Maxwellian ions. Figure~\ref{fig:pressures}a compares the measured IAW feature with forward Maxwellian calculations to use as reference. The overall broadening is consistent with $T_i^{(d)}\lesssim 500\,\rm{eV}$.

We show consistency with pressure balance by considering the energy Rankine-Hugoniot equation in the shock reference-frame. The plasma is isobaric around the shock with total pressure $P_{\rm total} = p_M + p_{\text{ram}} + p_{th,e} + p_{th,i}$, where $p_M = B^2/2\mu_0$ is the magnetic pressure, $p_{\text{ram}}=\rho \tilde{u}^2$ ($\tilde{u}= |u-v_{sh}|$) is the shock-frame ram pressure, and $p_{th,e}=k_B n_e T_e$ and $p_{th,i}=k_B n_i T_i$ are the electron and ion thermal pressures, respectively. Figure~\ref{fig:pressures}a shows the pressure components in the undisturbed upstream and the downstream. As expected, the upstream pressure is strongly ram-pressure-dominated, encompassing $\approx 96 \%$ of the total pressure. In the downstream, the flux-freeze condition implies that the magnetic field is correlated to the density, which entails that $p_M$ always plays a subdominant role in the energetics, as observed. The downstream is measured to have $p_{th,e} \approx 27\%$ and $p_{\text{ram}}\approx 27\%$. Hence, an $\approx 35\%$ deficit in the total pressure is found that can only be attributed to ion thermal pressure.  The ion temperature can then be inferred from (see \autoref{apx:energy})

\begin{equation}
T_i^{(d)} = \frac{\bar Z^{(d)}}{k_B n_e^{(d)}}
\Big(p_{\mathrm{tot}}^{(u)} - p_{\mathrm{ram}}^{(d)} - p_{\mathrm{th},e}^{(d)} - p_M^{(d)}\Big), 
\end{equation}

\noindent yielding $T_i^{(d)}=460\pm 170$ eV. 

As a sanity check, we compare the IAW data downstream of the shock with a forward calculation of the dynamic form factor $S(k, \omega)$, using ion parameters from the pressure balance analysis. \autoref{fig:pressures}b shows the IAW spectrum at $t= 8.2$ ns. The spectrum is skewed with a relatively prominent peak at around $527.5$ nm and seemingly smaller peaks towards bluer wavelengths. However, the signal-to-noise ratio is low enough that the intensity presents significant error bars and vertical scatter (the smaller peaks are not prominent enough to confidently fit them). On the figure, we present a forward calculation assuming a proton-dominated plasma ($15\%$ carbon ion abundance) with $T_i=450$ eV for both hydrogen and carbon ions. The observed overall broadening of the IAW is consistent with our estimated ion temperature for a small fraction of carbon. In \autoref{apx:energy} we show that carbon abundances do not play a significant role in the electron heating, even at $50\%$-$50\%$ proton-carbon mix. We note that since the ion velocity distribution function is not a Maxwellian, we do not expect a perfect fit, and carrying out a fit with a more sophisticated model is beyond the scope of this paper. Nevertheless, we can compare the data with a forward calculation to gain confidence in our inference.

Our pressure balance calculation, consistent with the broadening of the IAW spectrum, together with the downstream electron temperature measured directly, yields an electron ion temperature ratio $T_e^{(d)}/T_i^{(d)}= 0.8 \pm 0.3$. Below we discuss classical heating mechanisms (adiabatic compression and collisional coupling) and astrophysics relevance.


\subsection{Electron heating}

We now analyze electron heating processes in the shock.  As shown above, the electron heating is super-adiabatic, with $T_e^{(d)} / T_e^{(u)} > R^{2/3}$. We now show that it also cannot be explained by electron-ion collisions, which are sub-dominant but non-zero in these laboratory conditions.

\begin{figure}
    \centering
    \includegraphics[width=1\linewidth]{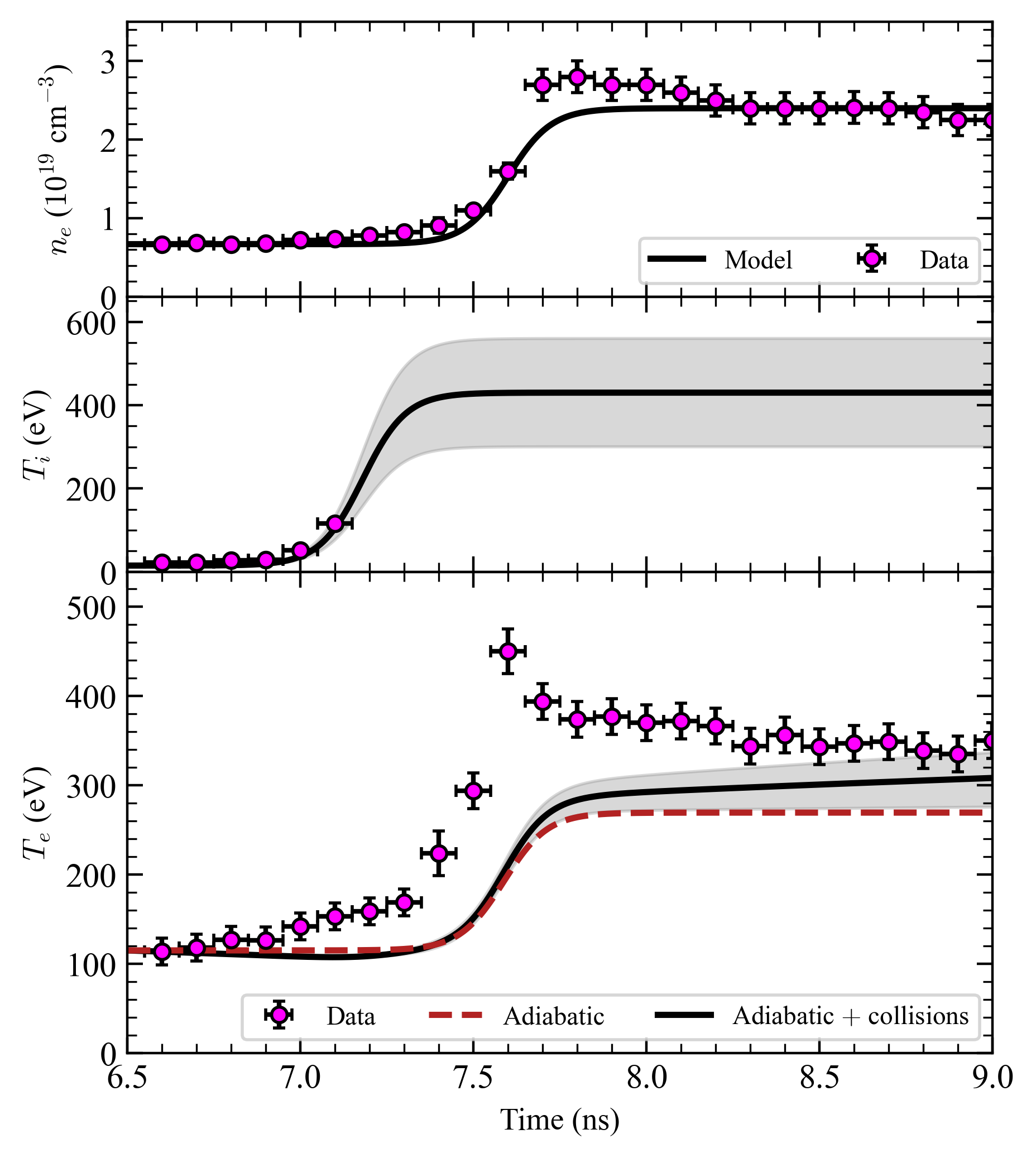}
    \caption{a) Parameterized electron density and data. b) Parameterized ion temperature and data. Grey shaded region denotes band of uncertainty on the downstream ion temperature. c) Comparison of measured electron temperature to adiabatic and collisional-adiabatic heating model. Grey band is calculated by considering extreme values of $T_i$ in the downstream.}
    \label{fig:heating}
\end{figure}

The electron heating equation due to adiabatic compression and electron-ion collisions is
%
\begin{equation}\label{eq:electron heating}
\frac{dT_e}{dt} =
\frac{2}{3}\frac{T_e}{n_e}\frac{dn_e}{dt}
+ \sum_s \lambda_s \frac{Z_s^2 f_s n_e}{\bar Z}\,
\frac{\left(T_i - T_e\right)}{T_e^{3/2}}.
\end{equation}
where the first term describes adiabatic heating and the second term describes heating due to electron-ion equilibration from collisions, summed over all ion species $s$ with relative fraction $f_s$.  Here, $\lambda_s = 1.8 \times 10^{-19}m_e^{1/2}\ln(\Lambda)/m_s$ for ion mass $m_s$. Equation (\ref{eq:electron heating}) can be solved in the collisionless limit (i.e., pure adiabatic heating) and semi-analytically when $T_i \gg T_e$ \citep{Ryutov2011,Ross2012,Valenzuela-Villaseca2024}, otherwise it must be treated numerically (see \autoref{apx:heating}). To do the calculation, $n_e = n_e(t)$ and $T_i=T_i(t)$ are modeled as step-like hyperbolic tangent profiles (\autoref{fig:heating}a,b). The gray bands around the parameterized $T_i$ show the maximum and minimum values allowed by pressure balance (\autoref{fig:pressures}). This range of ion temperatures is used to calculate the full range of plausible resulting electron temperatures $T_e(t)$ under this heating model. \autoref{fig:heating}c shows the results of the model in comparison to the observations, and
we observe several features.  First, we observe a super-adiabatic heating ahead of the shock, consistent with a foot region where energy can be deposited by reflected particles. Second, in the shock itself, adiabatic compression (red) heats the electrons up to at most $270\pm 20$ eV.
The addition of collisions (black) does provide some electron heating, but it is subdominant, adding $\approx 30$ eV, implying an experimentally observed $T_e^{(d)}$ excess. We check our calculation with a simple back-of-the-envelope estimate by adding the solution of the adiabatic heating equation and electron-ion collisional equilibration through the shock ramp, viz. $T_e^{(d)} \approx R^{2/3}T_e^{(u)} + (\tau_{\text{ramp}}/\tau_{eq}^{e\backslash i})(T_i^{(d)}-T_e^{(u)}) \approx 310$ eV ($\tau_{eq}^{e\backslash i}$ is the electron-ion equilibration time) for $\tau_{\text{ramp}}=0.25$ ns (see \autoref{apx:heating}). Even though this approximation overestimates the effect of electron heating, because it neglects that electrons and ions collisionally decouple at higher temperatures, it still underestimates the observed electron heating across the shock when compared to our measurements. Hence, the observed electron heating cannot be explained by adiabatic and collisional effects and must be collisionless in origin, perhaps mediated by wave-particle interactions.

\begin{table}
\small
\centering
\begin{tabular}{lcc}
\hline
\hline
                        &   Symbol          &   Experiment          \\
\hline
\textbf{Upstream}       &                   &                       \\                                
Magnetic field          & $B^{(u)}$         &	10.5 T              \\
Ion density             & $n_{i}^{(u)}$     &	$(6.7\pm0.5) \times 10^{18}$ cm$^{-3}$	\\
Electron temperature    & $T_{e}^{(u)}$     &	$115\pm15$ eV	    \\ 
Ion temperature         & $T_{i}^{(u)}$     &	$15\pm4$ eV			\\
Flow speed              & $u^{(u)}=v_{sh}$  &   $660\pm20$ km/s     \\
Ion inertial length     & $d_{i}^{(u)}$     &   $88\pm3$ $\mu$m      \\
Ion gyroperiod          & $\left[\omega_{ci}^{(u)}\right]^{-1}$ & $0.99$ ns  \\
Ion gyroradius          & $r_{i}^{(u)}$     &   $660\pm20$ $\mu$m          \\
\hline
\textbf{Downstream}     &                   &                       \\
Magnetic field          & $B^{(d)}$         &	$37.6\pm3.2$ T       \\
Ion density             & $n_{i}^{(d)}$     &	$(2.4\pm0.1) \times 10^{19}$ cm$^{-3}$	\\
Electron temperature    & $T_{e}^{(d)}$     &	$350\pm20$ eV		\\ 
Ion temperature         & $T_{i}^{(d)}$     &	$460\pm170$ eV		\\
Flow speed              & $u^{(d)}$         &   $475\pm30$ km/s      \\
Ion inertial length     & $d_{i}^{(d)}$     &   $47\pm1$ $\mu$m       \\
Ion gyroperiod          & $\left[\omega_{ci}^{(d)}\right]^{-1}$ & $0.28\pm0.02$  \\
Ion gyroradius          & $r_{i}^{(d)}$     &   $130\pm10$ $\mu$m           \\
\hline
\textbf{Dimensionless}  &                   &                        \\
Compression ratio       & $R$               &  $3.6\pm0.3$          \\
Temperature ratio       & $T_{e}^{(d)}/T_{i}^{(d)}$ &   $0.8\pm0.3$ \\
\hline
\hline
\end{tabular}
\caption{Plasma parameters in the OMEGA experiment.}
\label{tab:exp}
\end{table}

\begin{figure*}
    \centering
    \includegraphics[width=0.8\linewidth]{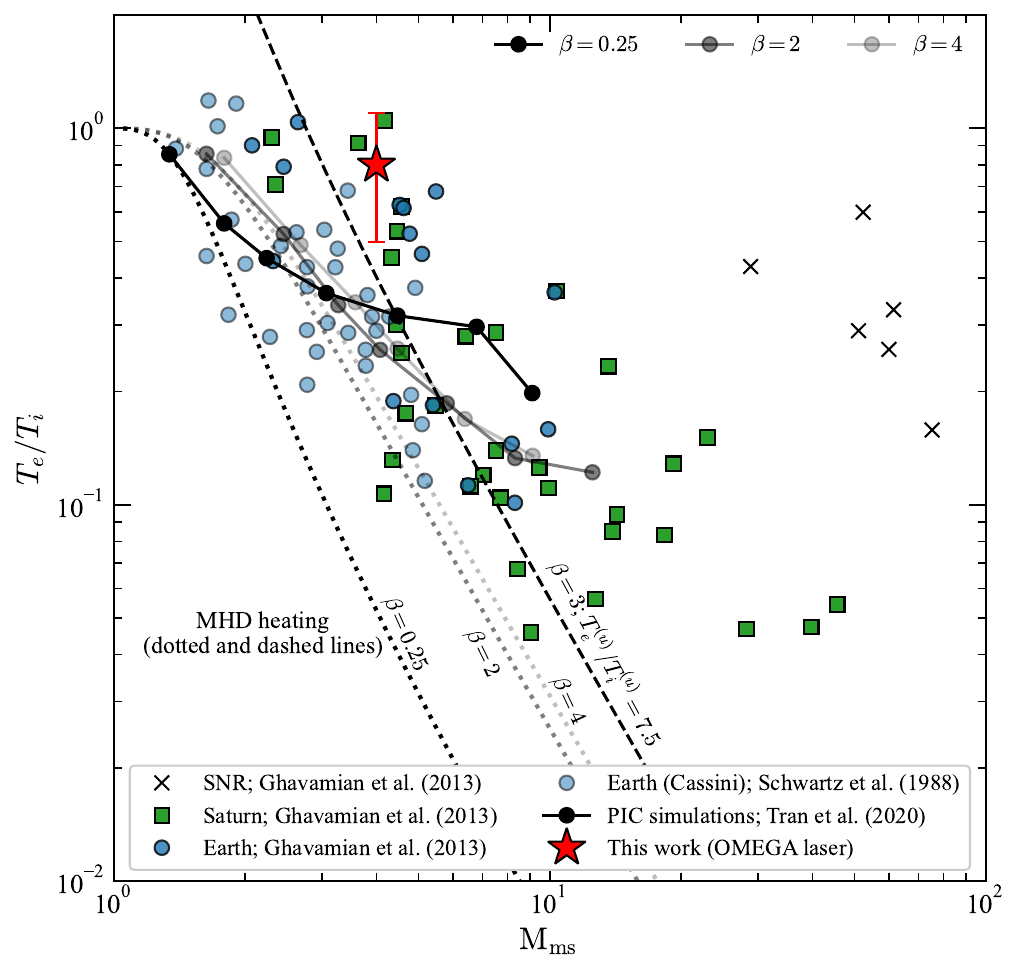}
    \caption{Scatter plot of downstream temperature ratio $T_e/T_i$ vs magnetosonic Mach number $\rm M_{ms}$ showing heliospheric and astrophysical shock measurements \citep{Schwartz1988,Ghavamian2013} together with the results from this work, marked with a red star. Also shown are MHD predictions for different upstream $\beta$, including the case $T_e^{(u)}=T_i^{(u)}$ (grayscale dotted line), the experimental condition $T_e^{(u)}/T_i^{(u)}=7.5$ (dashed black line), and a series of PIC simulations by \cite{Tran2020} (circle markers with solid line), with different upstream $\beta$ (grayscale).} 
    \label{fig:partition}
\end{figure*}

\subsection{Energy partition and comparison to MHD}

It is valuable compare our results to magnetohydrodynamics (MHD) theory, which is often used  in fluid and hybrid approaches for modeling SNR- and laboratory-relevant plasmas (e.g. \cite{Orlando2022,Orlando2023,Caprioli2014a,Caprioli2014b,Orusa2023,Orusa2025}). In the following, and for the sake of compactness, dimensionless parameters will be evaluated in the upstream, dropping the superscript as $\beta\equiv \beta^{(u)}$, $\mathrm{M}_{\mathrm{A}} \equiv \mathrm{M}_{\mathrm{A}}{}^{(u)}$, $\mathrm{M}_{\mathrm{s}} \equiv \mathrm{M}_{\mathrm{s}}{}^{(u)}$, and $\mathrm{M}_{\mathrm{ms}} \equiv \mathrm{M}_{\mathrm{ms}}{}^{(u)}$, but retaining it for temperatures and pressures. Typically, in MHD shocks it is assumed that electrons heat up adiabatically \citep{Burgess2015}, viz.
\begin{equation}\label{eq:adiabatic}
    T_e^{(d)}=R^{\gamma-1}T_e^{(u)},
\end{equation}
where $\gamma=5/3$ in our experiment and the ions take the rest of the internal energy (here we assume this energy is fully thermalized). The Rankine-Hugoniot momentum conservation of an MHD shock can be manipulated to find the total downstream-to-upstream pressure ratio
\begin{equation}\label{eq:pressure ratio}
\frac{p_{\rm th}^{(d)}}{p_{\rm th}^{(u)}} = 1 + \frac{2 \rm{M_A}^2}{\beta}\left( 1 - \frac{1}{R}\right)  +\frac{1}{\beta}\left(1-R^2\right).  
\end{equation}
Simultaneously the downstream-to-upstream pressure ratio satisfies $p_{\rm th}^{(d)}/p_{\rm th}^{(u)}= R~(T_e^{(d)}+T_i^{(d)})/(T_e^{(u)}+T_i^{(u)})$. Combining it with Equations~\ref{eq:adiabatic} and~\ref{eq:pressure ratio} yields the downstream ion temperature, viz.
\begin{equation}
    T_i^{(d)} = \frac{p_{\rm th}^{(d)}}{p_{\rm th}^{(u)}}\frac{\left(T_e^{(u)}+T_i^{(u)}\right)}{R}-R^{\gamma-1}T_e^{(u)},
\end{equation}
from which the downstream temperature ratio can be calculated from the compression ratio given by \citep{Burgess2015}
\begin{equation}
    R = \frac{2(\gamma+1)}{D + \sqrt{D^2 + 4(\gamma+1)(2-\gamma)M_A^{-2}}},
\end{equation}
where
\begin{equation}
D \equiv (\gamma-1) + \frac{2}{\rm M_s^2} + \frac{\gamma}{\rm M_A^2}.
\end{equation}
From the closed set of equations above, one can calculate the expected downstream electron to ion temperature ratio across the shock. \autoref{fig:partition} presents measurements of energy partition from SNR and heliophysical shocks together with our laboratory results. We also include PIC simulations from \cite{Tran2020} and equivalent MHD calculations for different values of $\beta$. In the simulations, it is typically assumed that $T_e=T_i$ upstream of the shock. Under this assumption, both MHD and kinetic PIC simulations predict a downstream ratio in the range $0.1$ to $0.3$, with the ratio decreasing further with increasing $\rm M_{ms}$. As discussed in the introduction, astrophysical and heliophysical measurements exhibit significant scatter of particle energy partition, often reaching higher values. In our experiments, we obtain $[T_e^{(d)}/T_i^{(d)}]_{\rm exp}=0.8\pm 0.3$, well above both MHD and PIC predictions. In particular, MHD predicts $[T_e/T_i]_{\rm MHD} = 0.2$. However, in our experiments electrons and ions are not equilibrated upstream of the shock, but rather $T_e^{(u)}=7.5~T_i^{(u)}$. Under these conditions, MHD predicts a higher ratio $[T_e/T_i]_{\rm MHD} = 0.38$ (see dashed line in \autoref{fig:partition}) but still underpredicts the measured value. The factor of $\approx2$ between the measured and predicted ratios shows that the anomalous, super-adiabatic heating observed in the laboratory modifies the downstream temperature ratio both by increasing $T_e$ and reducing the energy available for ion heating, thereby channeling a larger fraction of the dissipated energy into electrons and substantially altering the energy partition relative to the expectation from standard fluid closures.


\section{Conclusions} \label{sec:conclusion}

We reported the observation of fully developed, supercritical quasi-perpendicular collisionless shocks in the laboratory. Relevant parameters are summarized in \autoref{tab:exp}.  Despite being scaled to laboratory dimensions, the shocks have dimensionless parameters comparable to quasi-perpendicular shocks observed around the Earth or in supernova remnants. Our experiment shows that electrons are heated super-adiabatically (both ahead of the shock and in the downstream) even at moderate values of $\rm M_{ms}$, challenging a standard assumption in fluid and hybrid descriptions of collisionless shocks. We measure a downstream temperature ratio $T_e/T_i= 0.8 \pm 0.3$, consistent with spacecraft observations of planetary bow shocks by the Cassini, International Sun-Earth Explorer (ISSE), and Magnetospheric Multiscale (MMS) missions \citep{Schwartz1988,Ghavamian2013,Chen2018,Schwartz2022}, but outside the range of predicted values from numerical simulations and MHD theory (see \autoref{fig:partition}). 


More broadly, our measurements suggest that the downstream temperature ratio is not universal, but instead depends on the upstream partition of energy between electrons and ions. If the downstream ratio depends on upstream conditions, then it will be sensitive to the initial conditions adopted in numerical models, as well as to the local solar wind in the heliosphere and the interstellar medium in the vicinity of SNRs. Our results highlight the value of laboratory experiments as a means to point towards a productive interplay between experiments, kinetic simulations, spacecraft measurements, and astronomical observations.

An important next step is to test whether a fully kinetic PIC simulation initialized with upstream temperature ratio $T_e^{(u)}/T_i^{(u)}>1$ recovers the same downstream energy partition observed in our experiments. Such a study would provide a valuable way to identify the origin of the observed anomalous electron heating. Agreement with the experiment would show that the effect can emerge naturally from kinetic physics under the appropriate upstream conditions, whereas disagreement would point to missing ingredients that deserve further investigation. Therefore, the measurements presented in this paper provide a valuable absolute measure of energy partition for comparison to kinetic simulations of astrophysical plasmas, which predict a wide variety of heating mechanisms and resulting energy partition (e.g. \cite{Chen2018,Tran2020, Schwartz2022, Raymond2023, Morris2023, Vanthieghem2024,Lalti2024, Tran2024}). .


Future experiments will explore anisotropic heating with measurements along the magnetic field to help address the origin of this anomalous electron heating. Furthermore, these experiments open a path for future experimental campaigns to test theories of energy partitioning over a range of higher Mach numbers, especially around $\rm M_{ms}\sim10-20$ where simulations particularly struggle to reproduce observations of temperature ratios $T_e/T_i<0.1$ (see \autoref{fig:partition}).  The results presented here will also be used to study Thomson scattering in systems with non-Maxwellian VDFs. These efforts would allow direct measurements of the downstream ion energization. 


\acknowledgements
We are grateful to Ellie Tubman, Danny Russell, and Frederico Fiuza for insightful suggestions and energetic discussions. This work was supported by the U.S. Department of Energy National Nuclear Security Administration (NNSA) under Award No. DE-NA0004033. The experiment was conducted at the Omega Laser Facility with the beam time through the National Laser User's Facility (NLUF) user program. This material is based upon work supported by the Department of Energy National NNSA University of Rochester “National Inertial Confinement Fusion Program” under Award Number(s) DE-NA0003856 and the Department of Energy [National Nuclear Security Administration] University of Rochester “National Inertial Confinement Fusion Program” under Award Number(s) DE-NA0004144. This report was prepared as an account of work sponsored by an agency of the United States Government. Neither the United States Government nor any agency thereof, nor any of their employees, makes any warranty, express or implied, or assumes any legal liability or responsibility for the accuracy, completeness, or usefulness of any information, apparatus, product, or process disclosed, or represents that its use would not infringe privately owned rights. Reference herein to any specific commercial product, process, or service by trade name, trademark, manufacturer, or otherwise does not necessarily constitute or imply its endorsement, recommendation, or favoring by the United States Government or any agency thereof. The views and opinions of authors expressed herein do not necessarily state or reflect those of the United States Government or any agency thereof.

\facility{OMEGA-60 laser system, Omega Laser Facility, University of Rochester \citep{boehly+95}.}
\software{\texttt{PlasmaPy} \citep{plasmapy2024}.}

%






\appendix
\phantomsection
\label{sec:appendix}

\section{Optical Thomson Scattering Data Processing: \texttt{TSWiFT}} \label{apx:tswift}
The data from Omega was processed using the Thomson Scattering Work Pipeline and Fitting
Toolkit (\texttt{TSWiFT}), developed in-house at Princeton University. The \texttt{Jupyter}-based code processes \texttt{.h5} files provided by the facility, calibrates the CCD spectral images, takes time-averaged lineouts of
data, automatically corrects the background, and fits the Thomson spectra automatically using a
modified version of the \texttt{thomson.py} from the \texttt{Python} package \texttt{PlasmaPy} \citep{plasmapy2024}. Below, we sketch data pipeline and show how the main datasets in the manuscript were processed. The code documentation will be part of a separate publication (Valenzuela-Villaseca \& Schaeffer, in preparation).

\subsection{Inputs and calibration of CCD images}
\begin{figure}[h]
  \centering
  \includegraphics[width=9cm]{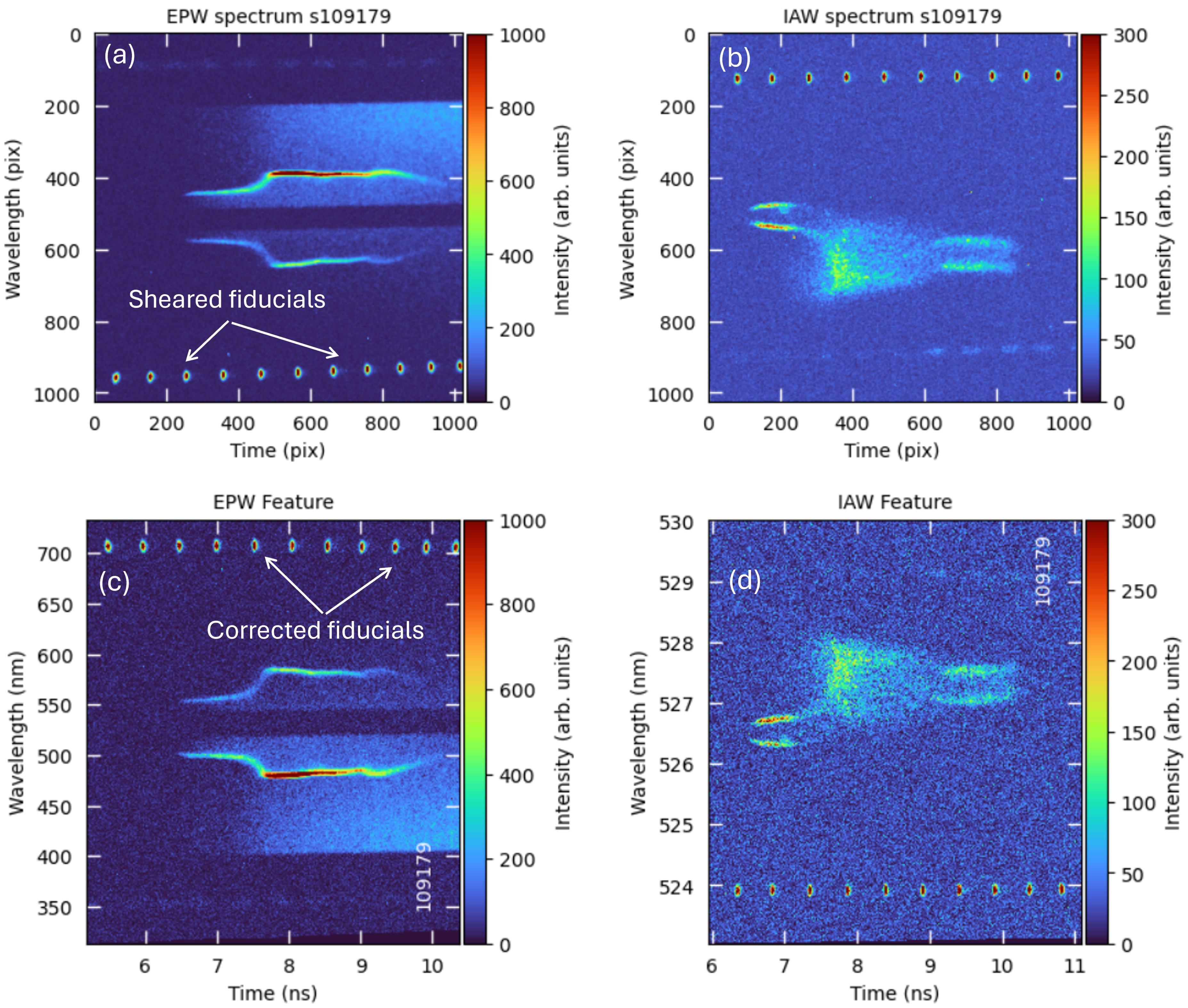}
  \caption{Example of CCD spectra. (a) Uncorrected EPW spectra. (b) Uncorrected IAW spectra.
  (c) Calibrated EPW spectra. (d) Calibrated IAW spectra.}
  \label{fig:A1}
\end{figure}

\texttt{TSWiFT} takes h5 files from the Omega laser facility containing the CCD spectral images and background (\autoref{fig:A1}a and b). In addition, the user can provide a .csv spreadsheet with shot parameters (such as time delays and pixel size), but these parameters can be entered directly in the \texttt{Jupyter} notebook.

The code automatically subtracts the dark current background, rotates the spectral images, plots the raw time-streaked CCD images from the electron-plasma wave (EPW) and ion-acoustic wave (IAW) channels. The code detects the facility timing fiducials inserted in the streak camera image and uses them to calibrate the time base. The user can fine-tune the overall timing to match the known probe beam timing. In addition, the code will correct the image shearing by fitting a line through the xy positions of the fiducials and applying a correction such that they lie at $y =$ constant. \texttt{TSWiFT} then uses a facility provided calibration from pixel to wavelength for the y-axis, assigning wavelength (\autoref{fig:A1}c and d).

\subsection{Spectral lineouts and background calibration}
The user can then take lineouts which average over a given timescale (in our analysis $=100$ ps, consistent with the resolution of the instrument). If multiple lineouts are taken, the code spaces them out equally with separation defined by the user. The user must introduce an overall spectral window (minimum and maximum wavelength to be considered in the lineout). In addition, the code also requests wavelengths outside the signal where it will be considered background. In general, the background noise on the IAW channel is flat, i.e. does not depend strongly on the photon energy, due to its narrow band. On the contrary, EPW usually exhibits strongly modulated background noise (see times $t > 8$ ns in \autoref{fig:A1}c for an example).

The code measures the average background intensity of the IAW in this region and subtracts it from the lineout. At the same time, it fits an arbitrary polynomial function to the EPW background (usually 4-th order is usually sufficient) and does the subtraction (\autoref{fig:A1}c). We emphasize that \emph{we do not fit a physical model to the background}, and therefore we do not draw physics conclusions from features such as the power balance between blue and red EPW resonances.

\begin{figure}[h]
  \centering
  \includegraphics[width=9cm]{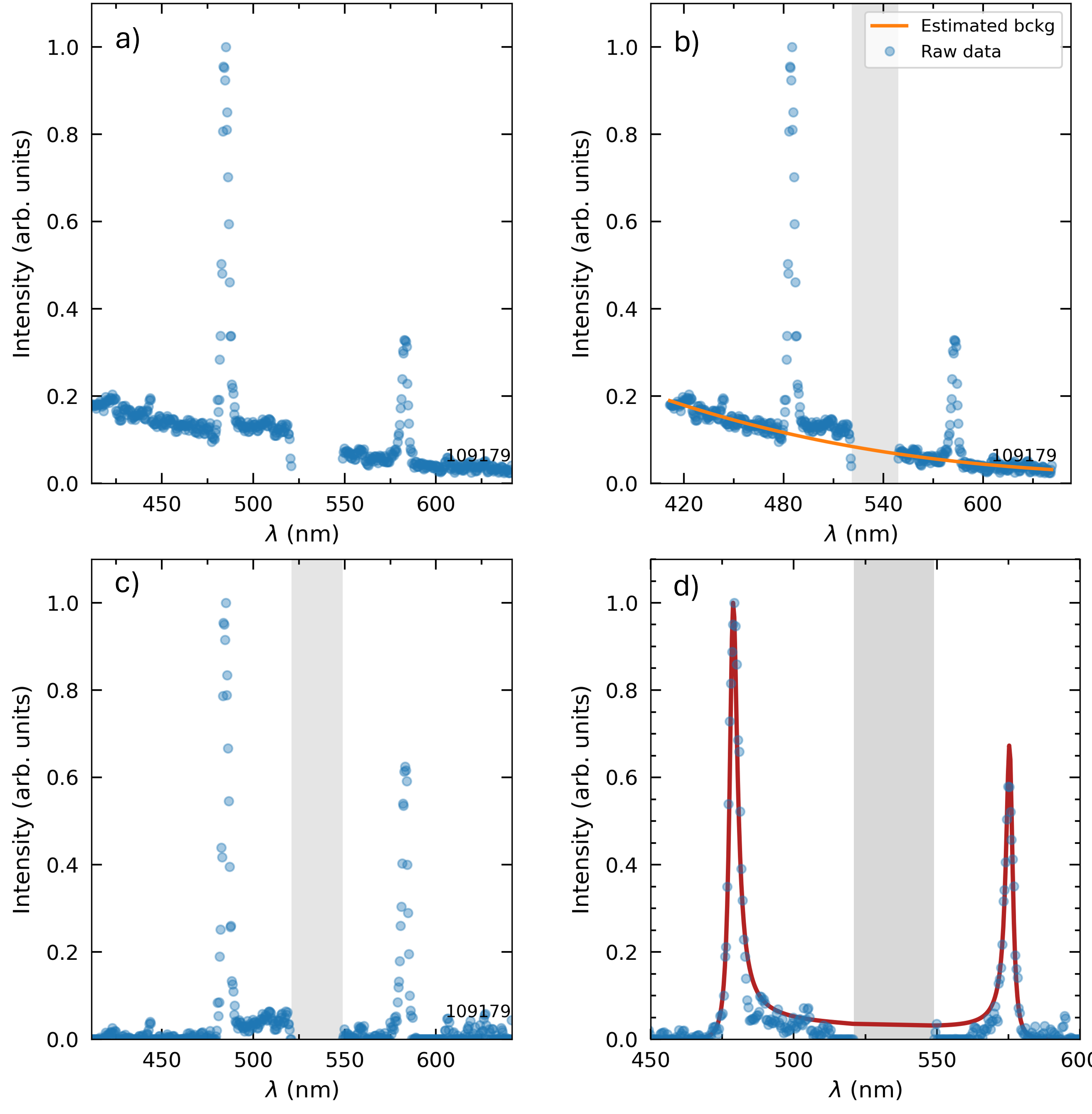}
  \caption{(a) Example of EPW spectral lineout. (b) Same as panel (a) but with automated background. (c) Background-corrected lineout. (d) Final fit.}
  \label{fig:A2}
\end{figure}

\subsection{Scattered power model}
The Thomson scattering power is
\begin{equation}
P_s \propto S(\mathbf{k},\omega)\left(1 + \frac{2\omega}{\omega_i}\right). 
\label{eq:A1}
\end{equation}
where $S(\mathbf{k},\omega)$ is the dynamical form factor shown in the main manuscript. The second term is a relativistic intensity modulation that is significant when electron-plasma waves have relativistic phase velocities. This both Doppler-shifts the laser-plasma interaction, narrows the blue light-cones, and broadens the red light-cones simultaneously \citep{Froula2011}. The net effect is that the blue resonance is enhanced compared to the red resonance. This correction is not implemented in the standard version of \texttt{PlasmaPy} but is important to fit the EPW data (the IAW feature is typically immune to it due to its narrow band). Therefore, \texttt{TSWiFT} has an extended package \texttt{thomson2.py} that includes the relativistic correction in the model.

\subsection{Fitting procedure}
\texttt{TSWiFT} utilizes the standard “differential method” implemented in \texttt{PlasmaPy}. The user inputs a model using \texttt{Python} dictionaries for each electron and ion species. The code convolves the OTS spectra with the instrument response, modelled with a gaussian profile. Forward modelling is available for an arbitrary number of electron and ion species, but a maximum of four electron populations and four ion populations can be fitted. In our experiments, we fit a single electron population and a maximum of two ion species. The user can set a given fit parameter to be fit or kept fixed.

\subsection{Output}
The code creates directories for the raw lineouts, estimated background, corrected lineouts, and fits. In each lineout, the code provides a report with best fit parameters. At the end of the pipeline, the code plots a summary of all spectra and their fits, similar to Figure 2 in the main body.

%
\section{Energy Conservation and Downstream Ion Temperature} \label{apx:energy}
We show how one can infer the downstream ion temperature via pressure balance across the discontinuity in the shock reference frame. Our analysis considers a multi-ion plasma, where ionic abundances are allowed to be different in the upstream and downstream. As we mentioned in the main body, the presence of multiple ion populations does not account for the observed electron heating, so we show the full calculation here. We first write down some mathematical relations between crucial plasma parameters, before parameterizing partial pressures, and then write down the equations used to infer the downstream ion temperature. In practice, we consider a two-ion plasma (hydrogen and carbon). Therefore, the carbon fraction is the only one of interest due to the normalization $\sum_\beta f_\beta =1$. At the end, we compare our inferred temperature with ion-acoustic wave (IAW) data from the experiment.

\subsection{Ion fractions}
In a given volume $V$, let the number of ions be

\begin{equation}
N_i = \sum_\beta N_\beta,
\label{eq:A2}
\end{equation}
where $N_\beta$ is the number of ions of species $\beta$. We define the ion fraction $f_\beta$ by dividing Eq.~\ref{eq:A2} by $N_i$, which yields
\begin{equation}
\sum_\beta N_\beta/N_i \equiv \sum_\beta f_\beta = 1.
\label{eq:A3}
\end{equation}

Thus, the overall ion number density $n_i$ is defined by dividing Eq.~\ref{eq:A2} by the volume, i.e.
\begin{equation}
n_i = \frac{N_i}{V}=\sum_\beta=\frac{N_\beta}{V}= \sum_\beta n_\beta,
\label{eq:A4}
\end{equation}
defining the ion number density $n_\beta$  of species $\beta$ also.

\subsection{Ion charge state, effective charge state, and quasi-neutrality}
Each ion species has a known charge state given by $Z_\beta$ via tabulated values and/or fitting of the Thomson scattering spectra. All ion species combined contribute to an effective charge state $\bar{Z}$ of the plasma given by the weighted average of individual ion charge states, with the weights given by the ion fractions defined in Eq.~\ref{eq:A3}, viz.
\begin{subequations}
\begin{align}
\bar Z &= \sum_\beta f_\beta Z_\beta, \label{eq:B5a} \\
       &= \sum_\beta \frac{n_\beta}{n_i}\, Z_\beta,  \\
       &= \frac{\sum_\beta n_\beta Z_\beta}{\sum_\beta n_\beta}, 
\end{align}
\end{subequations}
from which we obtain
\begin{align}
\bar Z \sum_\beta n_\beta &= \sum_\beta n_\beta Z_\beta = \bar Z\, n_i. 
\end{align}

Quasi-neutrality is therefore given by
\begin{align}
n_e &= \sum_\beta n_\beta Z_\beta = \bar Z \sum_\beta n_\beta = \bar Z n_i, 
\label{eq:B7}
\end{align}
where $n_e$ is the electron number density. Hence, the ion density for species $\beta$ is
\begin{align}
n_\beta &= f_\beta n_i = f_\beta \frac{n_e}{\bar Z}. 
\label{eq:B8}
\end{align}

\subsection{Pressure Components}
We work in the shock reference frame, so flow velocities are boosted ($\tilde{u}\rightarrow u$ from the main body). The strategy is to express the partial pressures as functions of known plasma parameters, such as density, temperature, and individual ion charge states. This will allow us to write the equations with the ion fractions as a free parameter which can be used to explicitly study the expected effect of different ion abundances in the plasma. The ram pressure is given by
\begin{align}
p_{\mathrm{ram}} &= \rho u^2
\end{align}
where $\rho$ is the mass density and $u$ is the plasma bulk velocity (for simplicity, electrons are assumed to be inertialess). Let us express $\rho$ in terms of $n_e$, viz.

\begin{subequations}
\begin{align}
\rho &= \sum_\beta m_\beta n_\beta \\
     &= \sum_\beta m_\beta f_\beta n_e \left(\sum_\alpha f_\alpha Z_\alpha\right)^{-1} \label{eq:B10b}\\
     &= n_e \sum_\beta \frac{f_\beta m_\beta}{\bar Z} \label{eq:B10c}
\end{align}
\end{subequations}
where we used Eqs.~\ref{eq:B7} and \ref{eq:B8} to obtain Eq.~\ref{eq:B10b}, we used Eq.~\ref{eq:B5a} to obtain Eq.~\ref{eq:B10c}, and $m_\beta$ is the ion mass of species $\beta$ given by the atomic weight $A_\beta$ and proton mass $m_p$ as $m_\beta=A_\beta m_p$ . Replacing the mass density, the ram pressure is
\begin{align} \label{eq:SM_ram pressure}
p_{\mathrm{ram}} &= n_e \left(\sum_\beta \frac{f_\beta m_\beta}{\bar Z}\right) u^2.
\end{align}

The total thermal pressure is given by
\begin{subequations}
\begin{align}
p_{\mathrm{th}} &= p_{\mathrm{th},e} + p_{\mathrm{th},i}, \\
                &= n_e k_B T_e + \sum_\beta n_\beta k_B T_\beta, 
\end{align}
\end{subequations}
where $p_{th,e}$ and $p_{th,i}$ are the electron and ion thermal partial pressures. We will take a conservative approach and reduce the degrees of freedom of the problem. Let us impose that all ions are in equilibrium, i.e. $T_\beta=T_i$ for all $\beta$. Hence, the total thermal pressure
\begin{subequations}
\begin{align}
p_{\mathrm{th}} &= n_e k_B T_e + k_B T_i \sum_\beta n_\beta, \\
                &= n_e k_B T_e + k_B T_i \sum_\beta \frac{f_\beta n_e}{\bar Z}, \\
                &= n_e k_B T_e + \frac{n_e k_B T_i}{\bar Z}\sum_\beta f_\beta, \\
                &= n_e k_B \left(T_e + \frac{T_i}{\bar Z}\right). 
\end{align}
\end{subequations}

Finally, the magnetic pressure does not depend on fluid properties and is given simply by
\begin{align}
p_M &= \frac{B^2}{2\mu_0}, 
\end{align}
where $B$ is the magnetic field and $\mu_0$ is the magnetic permeability of vacuum.

\begin{figure}[t]
  \centering
  \includegraphics[width=16cm]{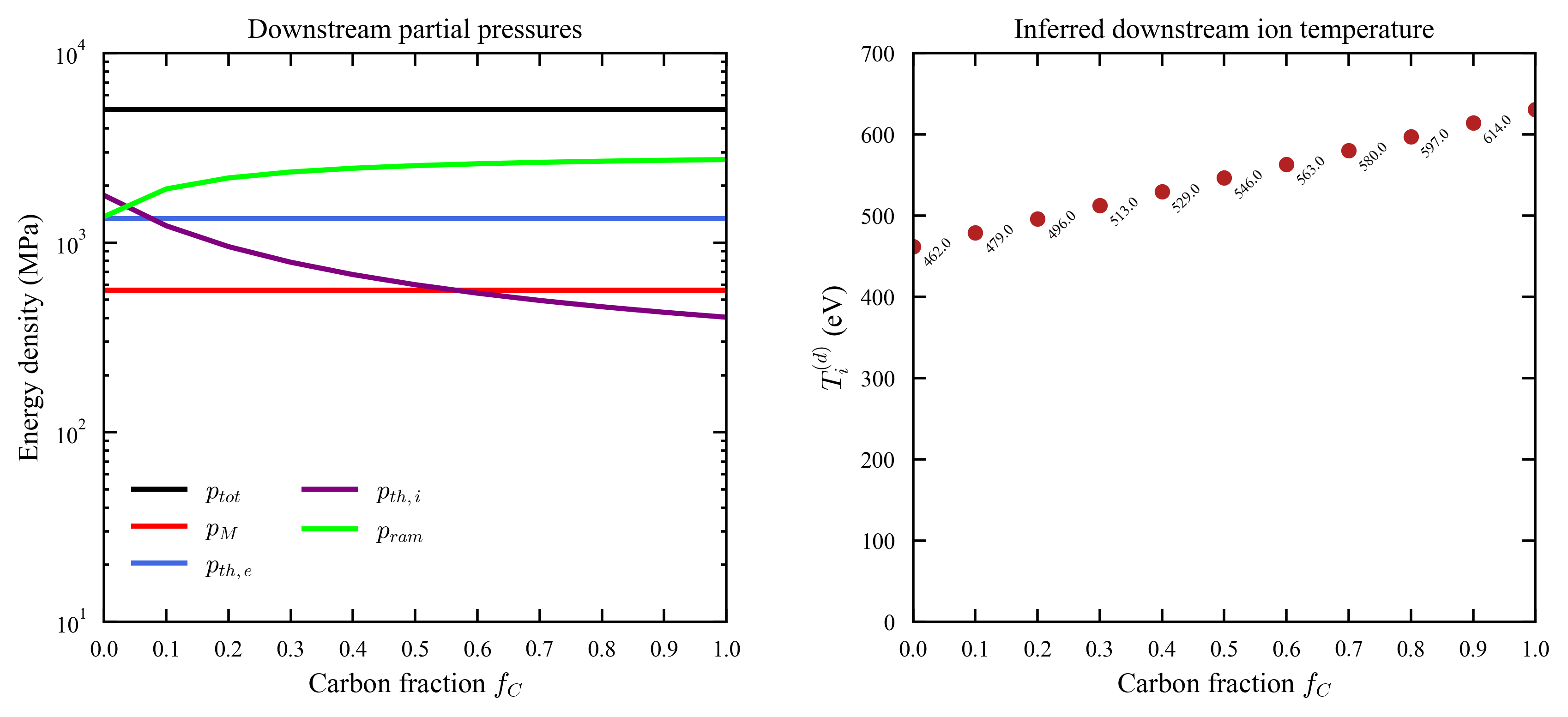}
  \caption{Downstream partial pressures (left) and inferred downstream ion temperature (right) as function of downstream carbon fraction.}
  \label{fig:B1}
\end{figure}

\subsection{Downstream ion temperature}
In the shock frame of reference, the total pressure $p_{tot}$ about the shock is constant and explicitly dependent on the ion fractions $\{f_\beta\}$ as
\begin{align}
p_{\mathrm{tot}} &= p_{\mathrm{ram}}(\{f_\beta\}) + p_{\mathrm{th}}(\{f_\beta\}) + p_M.
\end{align}

Experimentally, the total pressure is equal to the \emph{known} upstream pressure as measured using Thomson scattering, i.e. $p_{tot}=p_{tot}^{(u)}$; where superscripts $(u)$ and $(d)$ denote upstream- and downstream-evaluated parameters, respectively. (Not to be confused with the flow velocity $u$). Hence, the downstream ion temperature can be calculated as

\begin{align}
T_i^{(d)} &= \frac{\bar Z^{(d)}}{k_B n_e^{(d)}}
\Big(p_{\mathrm{tot}}^{(u)} - p_{\mathrm{ram}}^{(d)} - p_{\mathrm{th},e}^{(d)} - p_M^{(d)}\Big), 
\end{align}
or, explicitly
\begin{align}
T_i^{(d)} &= \frac{\bar Z^{(d)}}{k_B n_e^{(d)}}
\left(p_{\mathrm{tot}}^{(u)} - \frac{[B^{(d)}]^2}{2\mu_0}\right)
- \bar Z^{(d)} T_e^{(d)} - \sum_\beta \frac{A_\beta f_\beta^{(d)}}{k_B}m_p[u^{(d)}]^2.
\end{align}

It is useful to know how the partial pressures and inferred downstream ion temperature vary with different ion fractions. In our experiments, the species of interest are hydrogen and carbon which will henceforth denoted with $\{\beta\}=\{H,C\}$, respectively. The Thomson scattering data in the upstream indicates that it is a pure proton-electron plasma, viz. $f_H^{(u)}=1$ and $f_C^{(u)}=0$. Figure~\ref{fig:B1} (left panel) shows how the downstream partial pressures vary with carbon fraction. We find that as carbon populates the downstream, the ram pressure monotonically increases and therefore, to compensate, the ion thermal pressure must decrease. As a consequence, as presented in the right panel, the inferred ion temperature slightly increases with increasing carbon abundance towards $T_i^{(d)}\approx 500$ eV due to the scaling $p_{\rm{ram}}⁄p_{th,i}\sim \sum_\beta f_\beta A_\beta / \sum_\beta f_\beta Z_\beta \rightarrow 2 $ as $f_C\rightarrow 1$.

%
\section{Electron Heating Model} \label{apx:heating}

We now consider whether the presence of carbon in the downstream plasma could influence the inferred electron heating through the shock. Before addressing this quantitatively, we note that several features of the measurements rule out a significant role for electron heat transport or mixing from the piston plasma. The electron temperature is observed to increase continuously from the upstream region through the shock, with the onset of anomalous heating occurring upstream where the presence of carbon has been ruled out. Conversely, if heating or mixing from the piston was occurring, one would expect an monotonic increase of electron temperature through the downstream toward the piston region. This spatial signature is not observed. Quantitatively, this is consistent with the characteristic diffusion time from the contact discontinuity to the measurement region being much longer than the experimental timescale ($\approx 1$ ns), as using either Bohm diffusion, $D_{\rm B}=k_B T_e/(16eB)$, or perpendicular classical transport, $D_\perp \sim r_{g,e}^2 \nu_{ei}$, we estimate a cross-field heat diffusion time $\tau_{\rm{diff}}\sim L^2/D$ across the downstream of order $\sim 100$ ns \citep{Huba2016}. 

In this section we explicitly examine the effect of carbon abundance on the shock electron heating and show that its contribution is negligible. In the absence of heat losses (such as conduction and radiative cooling), the electron heating equation is given by a combination of adiabatic compression and collisional electron-ion thermal equilibration, i.e.
\begin{align}\label{eq:SM_electron heating}
n_e \frac{dT_e}{dt} &= A(t) + C(t)
\end{align}
where
\begin{align}
A(t) = (\gamma-1) T_e \frac{dn_e}{dt}&, \qquad
C(t) = \sum_\beta \nu^{\mathrm{eq}}_{e\beta}(T_\beta - T_e)
\end{align}
are adiabatic and collisional heating terms, respectively ($\gamma$ is the adiabatic index, assumed to be $=5⁄3$). The latter term depends on the electron-ion equilibration frequency \citep{Huba2016}
\begin{subequations}
\begin{align} 
\nu^{\mathrm{eq}}_{e\beta} &=
1.8\times10^{-19}\,
\frac{\left(m_e m_\beta\right)^{1/2} Z_\beta^2 n_\beta}{(m_\beta T_e + m_e T_\beta)^{3/2}}\ln\Lambda, \label{eq:C20a}\\ 
&\approx
1.8\times10^{-19}\,
\frac{m_e^{1/2}}{m_\beta}\frac{Z_\beta^2 n_\beta }{T_e^{3/2}}\ln\Lambda, \label{eq:C20b}
\end{align}
\end{subequations}
where, for simplicity we have adopted gaussian cgs units with temperatures measured in electronvolts. Equation~\ref{eq:C20b} is a good approximation as long as $T_\beta/T_i\ll m_\beta / m_e$ ($\approx 10^3$ for $\beta=H$, an electron-proton plasma). In Eqs.~C20, $\ln \Lambda\approx 8$ is the Coulomb logarithm. We define the coefficient
\begin{align}
\lambda_\beta &= 1.8\times10^{-19}\,\frac{m_e^{1/2}}{m_\beta}\,\ln\Lambda.
\end{align}

In Eq.~\ref{eq:C20b}, we can replace the ion density with the electron density through Eq.~\ref{eq:B8}. In addition, we assume a single ion temperature. The electron heating equation can be re-written as
\begin{align}\label{eq:SM_full_heating_equation}
\frac{dT_e}{dt} &=
\frac{2}{3}\frac{T_e}{n_e}\frac{dn_e}{dt}
+ \sum_\beta \lambda_\beta \frac{Z_\beta^2 f_\beta n_e}{\bar Z}\,
\frac{T_i - T_e}{T_e^{3/2}}.
\end{align}

To model the shock, we prescribe $n_e=n_e(t)$ and $T_i=T_i(t)$ as the step-like profiles (see \autoref{fig:B1})
\begin{subequations}
\begin{align}
n_e(t) &=
\frac{n_{\max}+n_0}{2}
+ \frac{n_{\max}-n_0}{2}\,
\tanh\!\left[\frac{2(t-t_{\mathrm{shock}})}{\tau_{\mathrm{ramp},n}}\right] \\
T_i(t) &=
\frac{T_{i,\max}+T_{i,0}}{2}
+ \frac{T_{i,\max}-T_{i,0}}{2}\,
\tanh\!\left[\frac{2(t-t_{\mathrm{shock}}-t_{\mathrm{foot}})}{\tau_{\mathrm{ramp},T_i}}\right] 
\end{align}
\end{subequations}
where $n_{max}$ and $n{_e} (0)$ are the peak and initial electron number density, $t_{\rm{shock}}$ is the time at which the shock passes through the probe, $t_{\rm{ramp},n_e}$ is the density jump ramp time, $T_{i,\rm{max}}$ and $T_i (0)$ are the peak and initial ion temperature, $t_{\rm{foot}}$ is the pre-delay of the foot ahead of the shock, and $t_{\rm{ramp}, T_i}$ is the ion temperature jump ramp time. We proceed to solve two approximate limits of Eq.~\ref{eq:SM_full_heating_equation} before the full equation numerically. In the main text we have used a full numerical solution, but it is instructive to support the
numerical results with the analytic calculations which we now present.

\subsection{Case I. Adiabatic heating}
In the absence of collisions in an ideal gas, equation \ref{eq:SM_electron heating} is reduced to a simple adiabatic heating equation 
\begin{align}
\frac{dT_e}{dt} &= \frac{2}{3}\frac{T_e}{n_e}\frac{dn_e}{dt} 
\end{align}
which is separable. By direct integration one finds
\begin{align} \label{eq:adiabatic heating}
T_e(t) &= \left(\frac{n_e(t)}{n_e(0)}\right)^{2/3} T_e(0). 
\end{align}

\subsection{Case II. Adiabatic and collisional heating in the $T_e \ll T_i$ limit}
Assuming a single ion species, defining $\lambda\equiv\lambda_\beta$, and dropping the last term of equation \ref{eq:SM_electron heating} one finds
\begin{align}\label{eq:S26}
\frac{dT_e}{dt} &= \frac{2}{3}\frac{T_e}{n_e}\frac{dn_e}{dt}
+ \lambda \frac{n_e}{T_e^{3/2}} T_i. 
\end{align}
We proceed to solve this equation exactly as \cite{Ryutov2011,Ross2012,Valenzuela-Villaseca2024}, we isolate the term with ion temperature on the right-hand-side and multiply both sides of equation \ref{eq:S26} by the integration factor $5T_e^{5/2}/2n_e^{2/3}$ such that the left-hand-side can be re-written as a perfect derivative

\begin{subequations}
\begin{align}
\frac{5}{2}\frac{T_e^{3/2}}{n_e^{5/3}} -
\frac{5}{3}T_e^{5/2}\frac{\dot{n_e}}{n^{8/3}}
&=\frac{5\lambda}{2}n_e^{-2/3}T_i, \label{eq:C27a}\\
\frac{d}{dt}\!\left(\frac{T_e^{5/2}}{n_e^{5/3}}\right)
&= \frac{5\lambda}{2}\, n_e^{-2/3} T_i \label{eq:C27b}
\end{align}
\end{subequations}
where, for compactness, we have written time derivatives using $\dot{a}\equiv da⁄dt$, in Eq.~\ref{eq:C27a}. Equation \ref{eq:C27b} can be integrated, yielding a semi-analytic Ryutov-Ross-like solution for the electron temperature
\begin{equation}
T_e(t) = \left[ \left(\frac{n_e(t)}{n_e(0)}\right)^{5/3} T_e(0)^{5/2} + \frac{5\lambda}{2}\, n_e(t)^{5/3} \int_0^{t} n_e(t')^{-2/3}\, T_i(t')\, dt' \right]^{2/5},\label{eq:RyutovRoss} 
\end{equation}
where the integral will depend on the parameterized profiles $n_e=n_e (t)$ and $T_i=T_i (t)$ (Eqs.~C23) and be evaluated analytically in some cases, and numerically in general. Notably, Eq.~\ref{eq:RyutovRoss} reduces to the adiabatic solution \ref{eq:adiabatic heating} in the collisionless limit $\lambda\rightarrow 0$.

It is useful to estimate the expected electron heating without integrating, numerically or otherwise. To first order, we can simplify the solution to equations \label{eq:full heating equation} and \ref{eq:S26} by adding together the adiabatic solution \ref{eq:adiabatic heating} with the collisional equilibration term multiplied by a characteristic timescale given by the ramp, i.e. estimate the downstream electron temperature as
\begin{equation}
T_e^{(d)} \approx \left(\frac{n_e^{(d)}}{n_e^{(u)}} \right)^{2/3} T_e^{(u)} +
\left(\frac{\tau_{\rm{ramp}}}{\tau_{eq}^{e\backslash i}} \right)\left(T_i^{(d)} - T_e^{(u)} \right),
\end{equation}
where $\tau_{\rm{ramp}}$ is the shock ramp time, and $\tau_{eq}^{e\backslash i}=[\nu_{eq}^{e\backslash i}]^{-1}=[T_e^{(u)} ]^{3⁄2}/\lambda \bar{Z}^{(d)} n_e^{(d)}$ is the electron-ion equilibration time. For an electron-proton plasma ($\bar{Z}=1$), and taking characteristic values $n_e^{(u)}=6.7\times 10^{18}$ cm$^{-3}$, $n_e^{(d)}=2.4\times 10^{19}$ cm$^{-3}$, $T_e^{(u)}=115$ eV, $\tau_{\rm{ramp}}=0.25$ ns, and $T_i^{(d)}=460$ eV, we estimate $\tau_{eq}^{e\backslash i}=2.25$ ns and a downstream electron temperature $T_e^{(d)} \approx 270$ eV$+40$ eV $\approx 310$ eV, which implies that electron-ion collisional coupling is negligible in the heating equation, and yields a downstream electron temperature significantly smaller than the experimentally measured value $T_{e,\rm{exp}}^{(d)}\approx 350 \pm 20$ eV. Notice that this calculation overestimates the final electron temperature, as it neglects that the electrons collisionally decouple from the ions as they heat up.

\subsection{Case III. Full heating equation}

\begin{figure}[h]
  \centering
  \includegraphics[width=10cm]{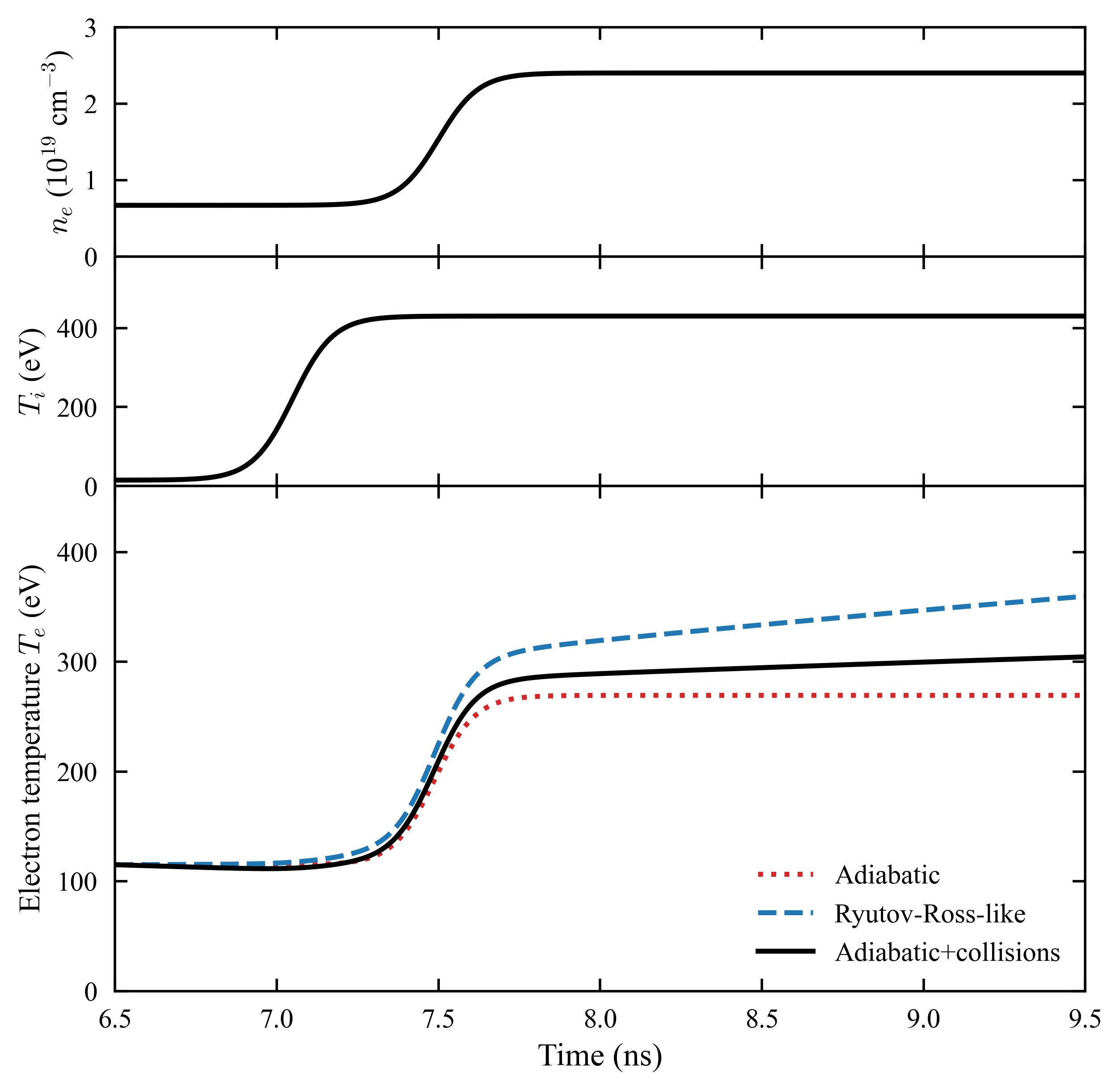}
  \caption{Calculated electron temperature evolution across the probe. Top: Prescribed electron number density profile. Middle: Prescribed ion temperature profile. Bottom: Calculated electron temperature evolution}
  \label{fig:C1}
\end{figure}

Given the prescribed density and ion temperature profiles (viz. Eqs.~C23), the full heating equation \ref{eq:SM_full_heating_equation} is solved numerically, resulting in \autoref{fig:C1}. The results confirm our first-order estimation. Electron-ion collisions marginally contribute to super-adiabatic heating (in \autoref{fig:C1}, the blue and black curves exceed the red dotted curve corresponding to adiabatic heating). However, on the timescales of interest, electron-ion collisions add $<40$ eV above the adiabatic temperature $T_{(e,\rm{adiabatic}}\approx 270$ eV (viz. equation \ref{eq:RyutovRoss}). Arguably, most importantly, it shows that the electron temperature saturates at $T_e \leq 300$ eV and does not significantly continue to heat up in the timescales of interest. We conclude that adiabatic compression together with electron-ion collisions are insufficient to explain the observed downstream electron temperature value $T_e=350\pm±20$ eV.

\subsection{Electron heating in a multi-ion component plasma}
\begin{figure}[h]
  \centering
  \includegraphics[width=12cm]{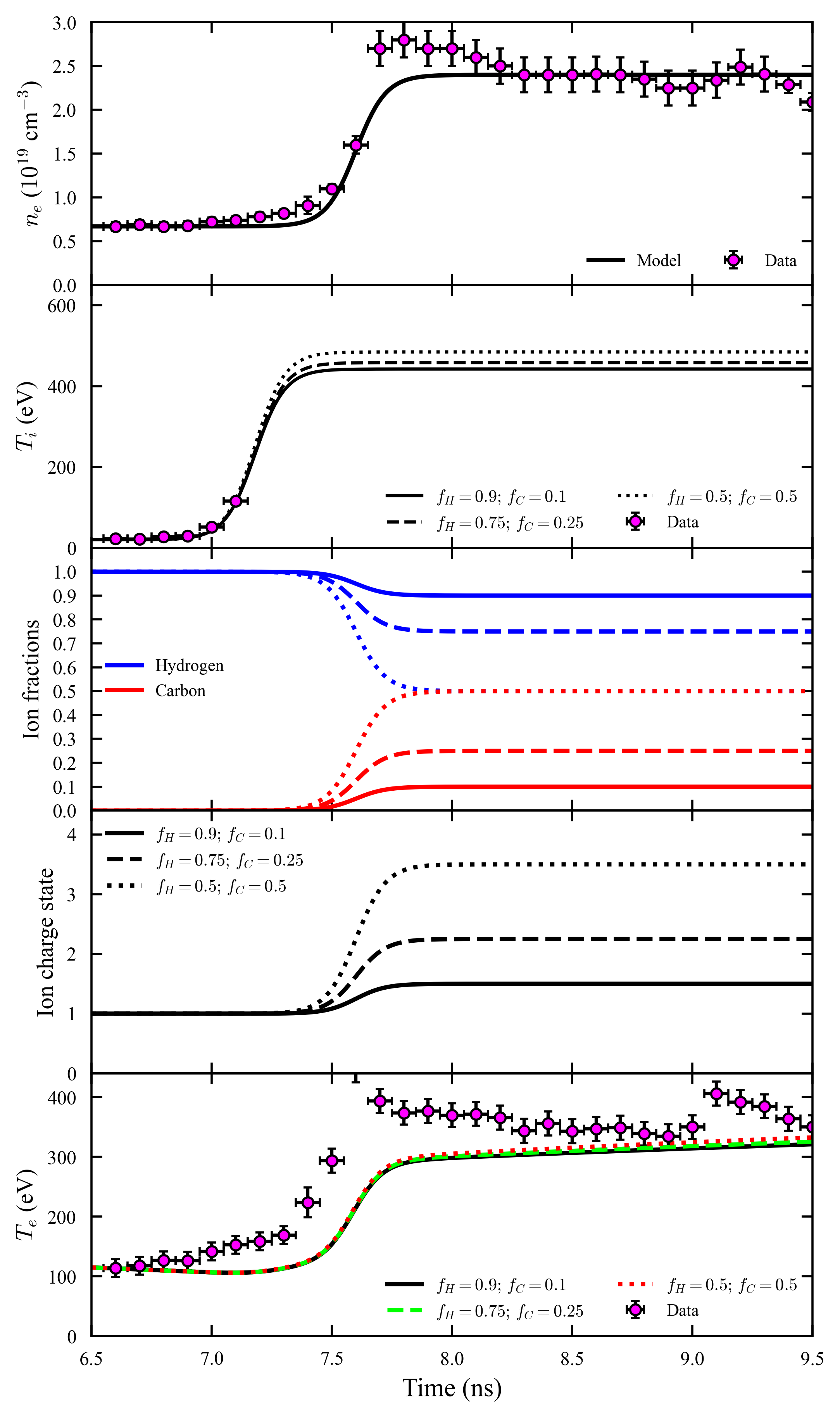}
  \caption{Calculated profiles and electron heating for different hydrogen/carbon fractions ($f_H$, $f_C$). Throughout the figure, solid lines denote $90\%$ hydrogen, $10\%$ carbon; dashed lines, $75\%$ hydrogen, $25\%$ carbon; dotted lines, $50\%$ hydrogen, $50\%$ carbon.}
  \label{fig:C2}
\end{figure}

\autoref{fig:C2} shows electron temperatures for different downstream proton and carbon abundances and ion temperatures consistent with Rankine-Hugoniot conditions. The effect of multi-ion components in the collisional heating is negligible

\bibliographystyle{aasjournal}
\bibliography{bibliography}



\end{document}